\documentclass[11pt]{article}
\usepackage{a4wide,graphicx,amsmath,amssymb,cite,longtable,multicol,array}
\usepackage{cancel}
\usepackage{comment}
\usepackage{multirow}

\newcommand{\lsim}{\raisebox{-0.13cm}{~\shortstack{$<$ \\[-0.07cm] $\sim$}}~}
\newcommand{\gsim}{\raisebox{-0.13cm}{~\shortstack{$>$ \\[-0.07cm] $\sim$}}~}

\newcommand{\nn}{\nonumber}
\newcommand{\bea}{\begin{eqnarray}}
\newcommand{\eea}{\end{eqnarray}}
\newcommand{\ba}{\begin{array}}
\newcommand{\ea}{\end{array}}

\newcommand{\ii}{{\rm i}}

\newcommand{\e}{{\rm e}}

\def\slash#1{\setbox0=\hbox{$#1$}%
  \rlap{\ifdim\wd0>.7em\kern.22\wd0\else\kern.1\wd0\fi /}#1}

\makeatletter
\@addtoreset{equation}{section}

\makeatother

\newcommand{\muegamma}{\mu\to\e\gamma}
\newcommand{\mueee}{\mu\to\e\e\overline{\e}}
\newcommand{\mueN}{\mu\,{\rm N\to\e\, N}}
\newcommand{\mue}{\mu-{\rm e}}

\def\Z{{\rm Z}}

\allowdisplaybreaks

\begin{document}

\hfill\begin{tabular}{r}
CAFPE-144/10 \\
UG-FT-274/10\\
\end{tabular}

\bigskip
\bigskip
\bigskip
\bigskip

\begin{center}

\begin{Large}
\textbf{Lepton flavor violation in the Simplest Little Higgs model}
\end{Large}

\bigskip
\bigskip

\begin{large}
Francisco~del~\'Aguila, 
Jos\'e~I.~Illana
and
Mark~D.~Jenkins
\end{large}

\bigskip

\begin{it}
CAFPE and Departamento de F{\'\i}sica Te\'orica y del Cosmos, \\[1ex]
    Universidad de Granada, E--18071 Granada, Spain
\end{it}

\bigskip

{\tt faguila@ugr.es}, {\tt jillana@ugr.es}, {\tt mjenk@ugr.es}
\end{center}

\bigskip
\bigskip

\begin{abstract}
The flavor sector of Little Higgs models based on product groups, notably the 
Littlest Higgs with T parity (LHT), has been extensively studied and some amount
of fine tuning was found to be required to meet the experimental constraints.
However, no such attention has been paid to other classes of models. Here we
analyze the phenomenology of flavor mixing in the lepton sector of a simple
group model, the Simplest Little Higgs (SLH). We obtain the Feynman rules of
the SLH in the 't Hooft-Feynman gauge up to the necessary order and calculate
the leading contributions to the rare processes $\muegamma$, $\mueee$ and 
$\mu-\e$ conversion in nuclei. We find results comparable to those of the LHT
model, because in both cases they arise at the one-loop level. These require the flavor
alignment of the Yukawa couplings of light and heavy leptons at the per
cent level or an effective scale of around 10 TeV.
\end{abstract}

\thispagestyle{empty}

\newpage
\tableofcontents

\newpage
\section{Introduction}

It is well known that the Standard Model (SM) Higgs mass receives
quadratically divergent radiative corrections dependent on the cutoff scale of
the model. Since electroweak precision data (EWPD) requires that the Higgs
mass be at most of the order of the electroweak scale \mbox{$v\simeq 246$
GeV}, naturalness arguments demand that this cutoff scale be near the TeV to
avoid fine tuning the model parameters (hierarchy problem). Hence, new physics
effects are expected before or at the TeV scale. Many beyond SM scenarios have
been proposed to solve this hierarchy problem, such as supersymmetry,
technicolor and extra-dimensions among others. However, the same EWPD and
flavor physics in general disfavor new particles at scales somewhat below 
$\sim $ 10 TeV giving rise to the so called \emph{little (flavor) hierarchy 
problem}.

Little Higgs (LH) models \cite{ArkaniHamed:2001ca,ArkaniHamed:2001nc,
Schmaltz:2005ky,Perelstein:2005ka} are also an attempt to solve the hierarchy
problem, bridging the gap to $\sim $ 10 TeV. This is done making the Higgs a
pseudo-Goldstone boson of a new approximate global symmetry broken at a scale
$f \sim 1$ TeV. The Higgs mass is protected from the one-loop quadratically
divergent corrections and leaves only a two-loop sensitivity to a 10 TeV
cutoff which is not considered unnatural. Beyond this scale an unspecified
ultraviolet completion is needed but this completion can now elude EWPD
and flavor constraints.

There are several possible realizations of LH models depending on how
the new symmetries are implemented.  Broadly speaking, they can be separated 
into two categories: those that have the SM gauge group emerging from the 
diagonal breaking of the product of several groups (for instance 
$(SU(2)\times U(1))^N$) and those where it emerges from the breaking of a larger
simple group (for example $SU(N)\times U(1)$). The former are called 
\emph{product group models} while the latter are termed \emph{simple group 
models}. There are some general features characteristic of each of the 
approaches \cite{Han:2005ru}. In the absence of further symmetries, product 
group models have additional free parameters from the gauge couplings of the 
different groups that diagonally break down to the SM gauge group, the SM Higgs
can be embedded in definite models into a single sigma-model multiplet and it is
possible to make the fermion sector relatively simple. On the other hand, 
simple group models have the enlarged gauge group couplings fixed by the SM 
values but they require, in specific models, at least two sigma-model 
multiplets. Moreover, fermion multiplets must be extended to transform under the
new gauge group.

At any rate, LH models introduce new particles with masses of order
$f$. However, EWPD generally requires $f > 4$ TeV 
\cite{Csaki:2002qg,Csaki:2003si,Han:2005dz,Chen:2006dy},
reintroducing the little hierarchy problem. In product group models, these 
constraints can be alleviated by introducing an additional discrete symmetry, 
T-parity \cite{Cheng:2003ju,Low:2004xc,Cheng:2004yc}. This symmetry exchanges 
the gauge groups, making almost all new particles T-odd and all SM particles 
T-even. In this way, dangerous tree-level couplings of light fields with only 
one heavy particle are forbidden. This avoids large contributions from higher 
dimensional operators obtained by integrating out the new heavy fields, thus 
relaxing the tension between EWPD and a lower $f$, which can be of the order of
the TeV in this case \cite{Hubisz:2005tx}. 
As a byproduct, the predicted (now relatively light) new particles could be
eventually observed at high-energy colliders \cite{Han:2003wu,Han:2005ru}. 
This justifies the attention which these models have received. In contrast, 
simple group models have no consistent way of introducing a similar mechanism 
\cite{Cheng:2004yc}, and thus they are somewhat disfavored.

However, one must also consider flavor constraints which are in general more 
stringent and translate into more restrictive, and also complementary, limits
on the model parameters (see \cite{Buras:2009if} for a review and further 
references). Flavor violating processes depend on
the new heavy scale as well as on the misalignment of SM and heavy flavors. 
Hence, the corresponding limits can be satisfied sending $f$ to a high 
enough value or aligning both sectors with a high enough precision. 
For instance, in the Littlest Higgs model with T-parity (LHT) present bounds 
on lepton flavor violating (LFV) processes require $f \gsim 10$ TeV or a 
misalignment of at most 1\% between the SM and the heavy fermion mass matrices 
\cite{Choudhury:2006sq,Blanke:2007db,delAguila:2008zu,delAguila:2010nv}. 
These limits are not stronger because T-parity forbids tree-level 
flavor changing $Z$ couplings. 
At any rate, flavor changing neutral currents (FCNC) 
involving only SM external fields are induced by the exchange of (heavy) T-odd 
particles at one loop, resulting in the previous bounds and 
reintroducing a {\em little} flavor hierarchy problem. 
Flavor violation in the quark sector has also been addressed in the literature, 
both in the Littlest Higgs with \cite{Blanke:2006eb,Goto:2008fj,Blanke:2009am} 
and without T parity \cite{Buras:2006wk}.

Once $f$ is of the order of several TeV, it is of the order of the scale implied
by the EWPD bounds on simple group models.  This means that simple group models
and the LHT would be on similar footing as long as flavor constraints on the
former models are not more stringent than in the LHT case. As we shall show in
the following, the LFV limits on the Simplest Little Higgs (SLH) model
\cite{Kaplan:2003uc,Schmaltz:2004de} are comparable to those on the LHT.  
This is so since in this simple group case the matter content of the model 
guarantees the absence of tree-level charged lepton FCNC, 
and the corresponding LFV processes are then one-loop suppressed.
In the SLH model the new global symmetry is $(SU(3)\times U(1))^2$, where only
the diagonal subgroup $SU(3)\times U(1)$ is gauged.  This gauge symmetry is
broken at the scale $f$ into the SM gauge group $SU(2)_L\times U(1)_Y$. 
Left-handed (right-handed) matter fields transform as $SU(3)$ triplets
(singlets), implying only the addition of heavy quasi-Dirac neutrinos to
complete the lepton multiplets \cite{delAguila:2005yi,delAguila:2007ap}. 
Hence, only neutrinos have tree-level FCNC, although mixing only light and
heavy neutrinos at the order considered, with no immediate observable effect 
\cite{delAguila:2008hw,delAguila:2009bb}. 

The quark sector in the SLH is more involved.  There are two ways of embedding
the SM quark doublets into the new SU(3) triplets.  In any case, the quark
Yukawa Lagrangians allow for mixing between heavy and light quarks of all three
families. Since in this paper we concentrate on the contributions to the basic
LFV processes $\muegamma$, $\mueee$ and $\mu$--e conversion in nuclei, the
quark sector is only relevant to the last process.  For simplicity, we will 
suppress all quark related mixings in this case to evaluate only the lepton 
mixing effects. 

We find that predictions for the SLH model are similar to those of the LHT.  
For instance, the present bound on $\mu\, {\rm Au}\to\e\, {\rm Au}$ requires 
$f \gsim 14\, (16)$ TeV for the anomaly-free (universal) quark embedding and 
natural values of the other model parameters in the SLH case, to be compared
with 10 TeV for the LHT model. Alternatively, the misalignment parameter 
must be $\sin2\theta\lsim 0.005\ (0.004)$ in the former case, and 
$\sin2\theta\lsim 0.01$ for the latter. The processes $\muegamma$ and $\mueee$
give comparable but less stringent limits. LFV $\tau$ decays will not be 
presented here, as they are less restrictive, since the branching ratios are 
suppressed by a factor ${\cal B}(\tau\to\ell\bar\nu_\ell\nu_\tau)\approx 0.2$ 
and the experimental limits are several orders of magnitude weaker.

LH models, as technicolor models in the past \cite{Dimopoulos:1980fj}, exhibit 
a naturalness {\em flavor problem}, more demanding in principle than the 
little hierarchy problem. Obviously, one can argue that they have 
significative regions of parameter space allowed by present flavor constraints, 
but these are characterised by a large scale or a small mixing angle, 
or a combination of them, 
when using a relevant physical parameterization. 
Also some observables can have cancelling contributions in other regions 
but these are relatively narrow and not common to all of them. 
They are prime candidates for explaining any future observation of (lepton) 
flavor violation beyond the SM. 
Thus, it is important not only to investigate the requirements for constructing 
realistic LH models, but to eventually interpret any possible 
departure from the SM predictions which may be observed at ongoing or planned
high precision flavor experiments like MEG \cite{Mori:2007zza,
Adam:2009ci} at PSI and PRISM/PRIME \cite{Kuno:2005mm,Pasternak:2010zz} at J-PARC. 
A general discussion of different scenarios, including the LHT model 
as well as supersymmetric and extra dimensional models, can be found in 
\cite{Altmannshofer:2009ne,Buras:2010wr}. 
The comparison of muon and $\tau$ branching fractions, when the latter are known with 
a better precision, should eventually help to discriminate between specific models 
\cite{Buras:2010cp}. 
If one just relies on processes involving only the first two families, 
models can be essentially classified depending on the order where $\mueee$ appears, 
{\it i.e.} tree level, one loop, or beyond, compared to $\muegamma$, which is 
always loop suppressed as required by gauge invariance.  
On the other hand, the use of the polarization of the initial lepton
provides further observables to determine the model structure, as recently discussed in the LHT case \cite{Goto:2010sn}.

Here we will present limits on the SLH model derived only from LFV effects, 
which are similar to those of the LHT, as already emphasized, 
because in both cases they are one-loop suppressed. 
In Section \ref{FR} we detail the calculation of the necessary Feynman rules 
in the 't Hooft-Feynman gauge to obtain the prediction of the LFV processes 
in the SLH for the first time. 
In Section \ref{LFVprocesses} we present the general structure of the LFV 
branching ratios and form factors and calculate the different contributions 
of the SLH. 
Section \ref{num} is dedicated to describing the numerical predictions of the 
model, while our conclusions are summarized in Section \ref{conclusions}.

\section{Feynman Rules for the SLH model}
\label{FR}

Our objective is to study lepton flavor violation in the SLH model.  The only new source of LFV in the SLH model is the misalignment of the SM down-type lepton mass matrix with the new heavy neutrino mass matrix.   Since we also neglect SM neutrino masses and mixing effects, this is in fact the only source of LFV.  This means that mixing matrices only appear in vertices that couple SM leptons to the new heavy neutrinos and, since our external particles are charged (e, $\mu$ and quarks), only charged currents can contribute to the flavor change.  An extension of the SLH model including the observed neutrino masses can be found in \cite{delAguila:2005yi}.

The SLH model is introduced in \cite{Kaplan:2003uc,Schmaltz:2004de} and the interaction Lagrangians are derived in \cite{Han:2005ru}.  However, some additional manipulation is required since we prefer to work in the 't Hooft-Feynman gauge.  This forces us to keep the would-be-Goldstone bosons (WBGB) explicit when obtaining the Feynman rules.  Also, the LFV matrix elements are of order $v^2/f^2$ in general so we have the additional complication of obtaining all the couplings up to the necessary order to guarantee the consistency of the calculations to this degree of precision. 

Our first step is to obtain all the physically relevant fields in the model.  When the fields are originally defined, they are the interaction states and usually are either not mass diagonal or have non canonical kinetic terms or both.  This is especially true when we take our expressions to order $v^2/f^2$.  Many of the fields would need no transformations and would directly be the physical fields if there were no electroweak symmetry breaking (EWSB).  In many cases the notation is inspired by this fact.  Also, as expected in a simple group model, several constants that in principle have no specific value end up being determined by the requirement that the model must contain the SM within it.

In the following we define our basic interaction fields and analyze in turn the different sections of the Lagrangian.

\subsection{Basic fields and expansions}

The SLH model is an $SU(3)_L \times U(1)_X$ gauge theory.  This gauge symmetry is a diagonal subgroup of a global $\big(SU(3)\times U(1)\big)_1\times \big(SU(3)\times U(1)\big)_2$ group.  The global symmetry is spontaneously broken to $\big(SU(2)\times U(1)\big)_1\times \big(SU(2)\times U(1)\big)_2$ and the gauge symmetry reduces to the SM gauge group $SU(2)_L \times U(1)_Y$.  We will write the covariant derivative for this gauge group in the following form:
\begin{equation}
D_\mu=\partial_\mu - \ii gA_\mu^a T_a + \ii g_xQ_xB_\mu^x\, ,\qquad g_x=\frac{gt_W}{\sqrt{1-t_W^2/3}}\,, \label{covar}
\end{equation}
with $t_W=s_W/c_W$ the tangent of the Weinberg angle $\theta_W$.
Writing the $SU(3)$ generators $T_a$ in the fundamental representation $(\mathbf{3})$, the $SU(3)$ part works out as follows:
\begin{equation}
A^aT_a=\frac{A^3}{2}\left(\begin{array}{ccc} 1 & 0 & 0 \\ 0 & -1 & 0 \\ 0 & 0 & 0 \end{array}\right)
+\frac{A^8}{2}\left(\begin{array}{ccc} 1 & 0 & 0 \\ 0 & 1 & 0 \\ 0 & 0 & -2 \end{array}\right)
+\frac{1}{\sqrt{2}}\left(\begin{array}{ccc} 0 & W^+ & Y^0 \\ W^- & 0 & X^- \\ {Y^0}^\dagger & X^+ & 0 \end{array}\right)\,.
\end{equation}
Here we have introduced the definitions of the gauge bosons $A^3$, $A^8$, $B^x$, $W^\pm$, $X^\pm$, $Y^0$ and ${Y^0}^\dag$.  Some of these will have to be rewritten in terms of the massive physical gauge fields later on.  The value of the gauge coupling $g_x$ ends up being set by the requirement that the photon couple to the electric charge.

The scalar sector of the SLH is a non-linear sigma model.  There are two scalar multiplets $\Phi_{1,2}$ transforming as $({\bf 3},{\bf 1})$ and $({\bf 1},{\bf 3})$ under $SU(3)_1\times SU(3)_2$, respectively (each then a $(\mathbf{3})$ representation of the $SU(3)$ gauge group with a $U(1)_X$ hypercharge of $-1/3$) that include 
the SM Higgs doublets as well as new Goldstone bosons.  They can be expressed as follows:
\begin{eqnarray}
\Phi_1&=&\exp\left(\frac{\ii\Theta'}{f}\right)\exp\left(\frac{\ii t_{\beta}\Theta}{f}\right)
\left(\begin{array}{c} 0 \\ 0 \\ fc_\beta \end{array}\right)\,,\\
\Phi_2&=&\exp\left(\frac{\ii\Theta'}{f}\right)\exp\left(-\frac{\ii\Theta}{t_{\beta}f}\right)
\left(\begin{array}{c} 0 \\ 0 \\ fs_\beta \end{array}\right)\,,
\end{eqnarray}
where
\begin{eqnarray}
\Theta=&\left(\begin{array}{ccc} 0 & 0 & h^0 \\ 0 & 0 & h^- \\ {h^0}^\dagger & h^+ & 0 \end{array}\right) +& \frac{\eta}{\sqrt{2}} \left(\begin{array}{ccc} \kappa & 0 & 0 \\ 0 & \kappa & 0 \\ 0 & 0 & 1 \end{array}\right)\,, \\
\Theta'=&\left(\begin{array}{ccc} 0 & 0 & y^0 \\ 0 & 0 & x^- \\ {y^0}^\dagger & x^+ & 0 \end{array}\right) +& \frac{z'}{\sqrt{2}} \left(\begin{array}{ccc} \kappa' & 0 & 0 \\ 0 & \kappa' & 0 \\ 0 & 0 & 1 \end{array}\right)\,,
\end{eqnarray}
with $t_\beta=s_\beta/c_\beta$ the ratio of the vacuum expectation values of the two Higgs triplets.
To introduce EWSB we will substitute $h^0 = (v+H)/\sqrt{2} - i\chi$ and $h^\pm = -\phi^\pm$.  Note that the term $\exp\left(\frac{\ii\Theta'}{f}\right)$ can be rotated away using a gauge transformation $SU(3)_L \times U(1)_X$ as would be the case if we wished to work in the unitary gauge.  The structure of the $\Theta$ and $\Theta'$ matrices is determined by the broken generators of the gauge symmetry.  One can add identity matrices from the $U(1)$ group which allows us to set the first two elements of the diagonal matrix independently from the third element ($\kappa$ and $\kappa'$).  Certain choices are convenient \cite{Han:2005ru} but they are not needed in our calculation since they only affect neutral Goldstone bosons.  The scalar triplets are then expanded up to any given order in $v/f$ (when we include EWSB) and expansions of up to order $v^4/f^4$ are required in some cases to obtain the necessary precision for the couplings.

To build the fermion sector of the model, the SM fermions have to be included in representations of the larger SLH gauge group.  The simplest way to do this is to embed the SM fermions into $SU(3)$ triplets. The fermion sector of the SLH model can then be broken down as follows.
\begin{itemize}
\item Each lepton family consists of an $SU(3)$ left-handed triplet (\textbf{3}) and 2 right-handed singlets (\textbf{1}).  There is no right-handed light neutrino:
\begin{equation}
L_m^T=(\nu_L,\ell_L,\ii N_L)_m\,, \quad \ell_{Rm}\,, \quad  N_{Rm}\,.
\end{equation}
\item The structure of the quark fields depends on the embedding we select. In the so called {\em universal embedding}, the quark sector is analogous to the lepton sector. The only difference is that we have three right-handed singlets:
\begin{equation}
Q_m^T=(u_L,d_L,\ii U_L)_m, \quad u_{Rm}\,, \quad d_{Rm}\,, \quad U_{Rm}\,.
\end{equation}
However, this universal fermion sector leads to $SU(3)$ and $U(1)_X$ gauge anomalies although these do not affect the SM gauge group.  Since the SLH model is an effective theory this is not necessarily a problem because one can add additional fermions at the cutoff scale of the theory to cancel the anomaly.  On the other hand, one can construct a quark sector that is directly anomaly-free with no additional degrees of freedom \cite{Kong:2003tf,Kong:2003vm}. This is known as the {\em anomaly-free embedding} and it requires that the first two families contain $SU(3)$ left-handed conjugate triplet representations ($\bar{\textbf{3}}$) and three right-handed singlets.  The third family is analogous to the lepton sector:
\begin{align}
&Q_1^T=(d_L,-u_L,\ii D_L)\,, \ \    d_{R}\,, \quad u_{R}\,, \quad D_{R}\,, \\
&Q_2^T=(s_L,-c_L,\ii S_L)\,, \quad  s_{R}\,, \quad c_{R}\,, \quad S_{R}\,, \\
&Q_3^T=(t_L,b_L,\ii T_L)\,,  \qquad t_{R}\,, \quad b_{R}\,, \quad T_{R}\,.
\end{align}
\end{itemize}
The gauge representations and hypercharges for the fermion sector for the different embeddings are summarized in table~\ref{tab:embeddings}.

\begin{table}
\begin{center}
\begin{tabular}{|c||c|c|c|c|c|c|c|}
\hline
\multicolumn{8}{|c|}{Universal embedding (U)} \\
\hline
Fermion & $Q_{1,2}$ & $Q_3$ & $u_{Rm},U_{Rm}$ & $d_{Rm}$ & $L_m$ & $N_{Rm}$ & $e_{Rm}$ \\
\hline
$Q_x$ charge & $1/3$ & $1/3$ & $2/3$ & $-1/3$ & $-1/3$ & $0$ & $-1$ \\
\hline
$SU(3)$ rep. & $\mathbf{3}$ & $\mathbf{3}$ &$\mathbf{1}$ & $\mathbf{1}$ & $\mathbf{3}$ &$\mathbf{1}$ & $\mathbf{1}$\\
\hline \hline
\multicolumn{8}{|c|}{Anomaly-free embedding (AF)} \\
\hline
Fermion & $Q_{1,2}$ & $Q_3$ & $u_{Rm},T_{Rm}$ & $d_{Rm},D_{Rm},S_{Rm}$ & $L_m$ & $N_{Rm}$ & $e_{Rm}$ \\
\hline
$Q_x$ charge & $0$ & $1/3$ & $2/3$ & $-1/3$ & $-1/3$ & $0$ & $-1$ \\
\hline
$SU(3)$ rep. & $\bar{\mathbf{3}}$ & $\mathbf{3}$ &$\mathbf{1}$ & $\mathbf{1}$ & $\mathbf{3}$ &$\mathbf{1}$ & $\mathbf{1}$\\
\hline
\end{tabular}
\end{center}
\caption{Quantum numbers of the fermion fields for the universal and anomaly-free embeddings.\label{tab:embeddings}}
\end{table}

\subsection{Gauge and Goldstone boson sector}

In this section we analyze the pieces of the Lagrangian that involve only gauge and Goldstone bosons.

\subsubsection{Scalar Lagrangian}

From the gauge invariant Lagrangian
\begin{equation}
\mathcal{L}_\Phi = |D_\mu\Phi_1 |^2 + |D_\mu \Phi_2 |^2
\label{lagphi}
\end{equation}
we can readily obtain the charged gauge boson mass terms.  As a first approximation we keep terms up to order $v^2/f^2$.  To this order the charged boson sector is diagonal,
\begin{equation}
\mathcal{L}_\Phi \supset M_W^2W^+_\mu W^{-\mu} + M_X^2 X_\mu^+ X^{-\mu}\,,
\end{equation}
where
\begin{eqnarray}
M_W &=& \frac{gv}{2}\,, \\
M_X &=& \frac{gf}{\sqrt{2}}\left(1-\frac{v^2}{4f^2}\right)\,.
\end{eqnarray}
This level of precision is sufficient everywhere except when obtaining the correct $\mathcal{O}(v^2/f^2)$ couplings for Goldstone bosons.  We need to go up to order $v^4/f^4$ to obtain the corrections to the $W$ mass and higher order corrections to the gauge boson eigenstates for these couplings.  Taking our expansion up to order four, the mass terms then read:
\begin{align}
\mathcal{L}_\Phi \supset&\ \frac{g^2v^2}{4}\left[1-\frac{v^2}{6f^2}\left(\frac{c_\beta^4}{s_\beta^2}+\frac{s_\beta^4}{c_\beta^2}\right)\right]W^+_\mu W^{-\mu}
+ \frac{g^2f^2}{2}\left[1-\frac{v^2}{2f^2}+\frac{v^4}{12f^4}\left(\frac{c_\beta^4}{s_\beta^2}+\frac{s_\beta^4}{c_\beta^2}\right)\right]X^+_\mu X^{-\mu} \nn\\
&+ \left[  \frac{\ii v^3}{6\sqrt{2}f^3}\left(\frac{c_\beta^3}{s_\beta}-\frac{s_\beta^3}{c_\beta}\right) W^-_\mu X^{+\mu} + \textrm{h.c.} \right]\,.
\end{align}
To obtain the correct physical states we must rotate the original fields as follows:
\begin{eqnarray}
W^\pm & \rightarrow & W^\pm \pm \frac{\ii v^3}{3\sqrt{2}f^3}\left(\frac{c_\beta^3}{s_\beta}-\frac{s_\beta^3}{c_\beta}\right) X^\pm\,, \nonumber \\
X^\pm & \rightarrow & X^\pm \pm \frac{\ii v^3}{3\sqrt{2}f^3}\left(\frac{c_\beta^3}{s_\beta}-\frac{s_\beta^3}{c_\beta}\right)W^\pm\,. \label{gaugerot}
\end{eqnarray}
Note that the physical states $W$ and $X$ differ from the interaction states only by a term of order $v^3/f^3$.  This difference is irrelevant almost everywhere in our calculation, but is important in determining the would-be-Goldstone boson states.  The interaction fields $W$ and $X$ will be considered equal to the physical fields elsewhere.

The masses of the physical fields work out as:
\begin{eqnarray}
M_W & = & \frac{gv}{2}\left[1-\frac{v^2}{12f^2}\left(\frac{c_\beta^4}{s_\beta^2}+\frac{s_\beta^4}{c_\beta^2}\right)\right]\,,\\
M_X & = & \frac{gf}{\sqrt{2}}\left[1-\frac{v^2}{4f^2}+\frac{v^4}{24f^4}\left(\frac{c_\beta^4}{s_\beta^2}+\frac{s_\beta^4}{c_\beta^2}\right)\right] \simeq \frac{gf}{\sqrt{2}}\left(1-\frac{v^2}{4f^2}\right)\,.
\end{eqnarray}
The $\mathcal{O}(v^4/f^4)$ correction to $M_X$ is neglected since it is unimportant for our calculation.

The neutral sector is already non-diagonal at order $\mathcal{O}(v^2/f^2)$ and requires some more work:
\begin{equation}
\mathcal{L}_\Phi \supset M_{Y}^2 Y^{0\mu} Y^{0\dagger}_\mu + \left(\begin{array}{ccc} A_3&A_8&B_x \end{array}\right) \mathcal{M} \left(\begin{array}{c} A_3 \\ A_8 \\ B_x \end{array} \right)\,,
\end{equation}
\begin{equation}
\mathcal{M}=f^2\left(
\begin{array}{ccc} 
\frac{g^2v^2}{8f^2} & \frac{g^2v^2}{8\sqrt{3}f^2} & \frac{gg_xv^2}{12f^2} \\ 
\frac{g^2v^2}{8\sqrt{3}f^2} & \frac{g^2}{3}-\frac{g^2v^2}{8f^2} & -\frac{gg_x}{3\sqrt{3}}+\frac{gg_xv^2}{4\sqrt{3}f^2} \\
\frac{gg_xv^2}{12f^2} & -\frac{gg_x}{3\sqrt{3}}+\frac{gg_xv^2}{4\sqrt{3}f^2} & \frac{g_x^2}{9} \end{array}\right)\,.
\end{equation}
Diagonalizing this matrix, the masses at order $\mathcal{O}(v^2/f^2)$ are:
\begin{eqnarray}
\mathcal{L}_\Phi &\supset & \frac{1}{2} M_{Z'}^2 Z'^\mu Z'_\mu + \frac{1}{2}M_{Z}^2 {Z}^\mu {Z}_\mu + \frac{1}{2} M_{A}^2 A^\mu A_\mu + M_{Y}^2 Y^{0\mu} Y^{0\dagger}_\mu,\\
M_A&=&0\,,\\
M_{Z}&=&\frac{gv}{2c_W}\,, \\
M_{Z'}&=&\frac{\sqrt{2}gf}{\sqrt{3-t_W^2}}\left(1-\frac{3-t_W^2}{c_W^2}\frac{v^2}{16f^2}\right)\,,\\
M_{Y}&=&\frac{gf}{\sqrt{2}}\,.
\end{eqnarray}
The \emph{first order} mixing matrix for gauge bosons is then:
\begin{equation}
\left(\begin{array}{c} A^3 \\ A^8 \\ B_x \end{array}\right) = 
\left(\begin{array}{ccc} 
0 & c_W & -s_W \\
\frac{1}{\sqrt{3}}\sqrt{3-t_W^2} & \frac{s_W^2}{\sqrt{3}c_W} & \frac{s_W}{\sqrt{3}} \\
-\frac{t_W}{\sqrt{3}} & \frac{s_W}{\sqrt{3}}\sqrt{3-t_W^2} & \frac{c_W}{\sqrt{3}} \sqrt{3-t_W^2} \end{array}\right)
\left(\begin{array}{c} Z' \\ Z \\ A \end{array}\right)\,.
\label{vecmix}
\end{equation}
Additionally, the physical $Z$ and $Z'$ states also require the replacements:
\begin{eqnarray}
Z'&\to&Z'+\delta_Z Z \, , \nn \\
Z&\to&Z-\delta_Z Z' \, ,
\label{zmix}
\end{eqnarray}
where
\begin{equation}
 \delta_Z = -\frac{(1-t_W^2)\sqrt{3-t_W^2}}{8c_W}\frac{v^2}{f^2}\,.
\end{equation}

We now need to find the actual Goldstone eigenstates.  The ones that appear in the original expansion have non-diagonal kinetic terms to order $v^2/f^2$ and there is mixing of these states with the gauge bosons through terms $V^\mu\partial_\mu \phi$.  We need only the charged sector since the neutral Goldstone bosons do not contribute to lepton flavor mixing processes.

The kinetic terms for the charged Goldstone bosons and the Goldstone-gauge mixing terms read (interaction fields):
\begin{eqnarray}
\mathcal{L}_\Phi & \supset & \left[1-\frac{v^2}{6f^2}\left(\frac{c_\beta^4}{s_\beta^2}+\frac{s_\beta^4}{c_\beta^2}\right)\right]\partial_\mu \phi^+ \partial^\mu \phi^- 
+ \left[ 1-\frac{v^2}{2f^2}\right] \partial_\mu x^+ \partial^\mu x^- \nn\\ 
&& - \frac{v^2}{3f^2}\left(\frac{c_\beta^3}{s_\beta}-\frac{s_\beta^3}{c_\beta}\right)
(\partial_\mu \phi^+ \partial^\mu x^- +\partial_\mu \phi^- \partial^\mu x^+)\,,
\end{eqnarray}
\begin{equation}
\begin{split}
\mathcal{L}_\Phi \supset & \;  \ii W^-_\mu \frac{gv}{2}\left( \left(1-\frac{v^2}{6f^2} \left(\frac{c_\beta^4}{s_\beta^2}+\frac{s_\beta^4}{c_\beta^2}\right)\right)\partial^\mu \phi^+ - 
\frac{v^2}{3f^2}\left(\frac{c_\beta^3}{s_\beta}-\frac{s_\beta^3}{c_\beta}\right) \; \partial^\mu x^+\right) \\
& + X^-_\mu \frac{gf}{\sqrt{2}} \left(
\frac{v^2}{3f^2}\left(\frac{c_\beta^3}{s_\beta}-\frac{s_\beta^3}{c_\beta}\right) \; \partial^\mu \phi^+ - \left(1-\frac{v^2}{2f^2}\right) \partial^\mu x^+\right) + \textrm{h.c.}\, .
\end{split}
\end{equation}
We can combine all the necessary Goldstone transformations into a single pair of equations that express the original interaction eigenstates in terms of the final Goldstone states (to order $v^2/f^2$) associated to the $W$ and $X$ bosons and have canonically normalized kinetic terms:
\begin{eqnarray}
x^\pm & \rightarrow &-\left(1+\frac{v^2}{4f^2}\right)x^\pm \mp \ii\frac{v^2}{3f^2}\left(\frac{c_\beta^3}{s_\beta}-\frac{s_\beta^3}{c_\beta}\right) \phi^\pm\,, \nonumber \\
\phi^\pm & \rightarrow & \mp \ii\left(1+\frac{v^2}{12f^2}\left(\frac{c_\beta^4}{s_\beta^2}+\frac{s_\beta^4}{c_\beta^2}\right)\right)\phi^\pm\,. \label{goldrot}
\end{eqnarray}
The calculation of these states required the use of relation (\ref{gaugerot}) to obtain the $v^2/f^2$ corrections.

Taking both (\ref{goldrot}) and the gauge boson rotations (\ref{vecmix}) and (\ref{zmix}) into account, the relevant Feynman rules can now be obtained.  The results to order $\mathcal{O}(v^2/f^2)$ are given in table~\ref{tab:SVV.VSS}.

\begin{table}
\begin{center}
\begin{tabular}{|c|c|}
\hline
SVV & $K$ 
\\
\hline
$x^\pm X^\mp \gamma$ & $\pm \ii M_X$ 
\\
\hline
$\phi^\pm W^\mp \gamma$ & $\pm \ii M_W$
\\
\hline
$x^\pm X^\mp Z$ & $\mp \ii M_X \frac{c_W^2-s_W^2}{2c_Ws_W} \mp \ii \delta_\Z \frac{M_X}{2s_Wc_W^2\sqrt{3-t_W^2}}$
\\
\hline
$\phi^\pm W^\mp Z$ & $\pm \ii M_W t_W \pm \ii \delta_Z M_W\frac{1-t_W^2}{s_W\sqrt{3-t_W^2}}$
\\
\hline
$x^\pm X^\mp Z'$ & $\mp \ii \frac{M_X}{2s_Wc_W^2\sqrt{3-t_W^2}} \pm \ii \delta_Z M_X \frac{c_W^2-s_W^2}{2c_Ws_W}$
\\
\hline
$\phi^\pm W^\mp Z'$ &
$\pm \ii M_W\frac{1-t_W^2}{s_W\sqrt{3-t_W^2}}\mp \ii \delta_Z M_W t_W $ 
\\
\hline
\end{tabular}
\begin{tabular}{|c|c|}
\hline
VSS & $G$ 
\\
\hline
$\gamma x^\pm x^\mp$ & $\mp 1$
\\
\hline
$\gamma \phi^\pm \phi^\mp$ & $\mp 1$ 
\\
\hline
$Z x^\pm x^\mp$ & $\pm \frac{c_W^2-s_W^2}{2s_Wc_W}\mp \delta_Z\frac{1-t_W^2}{2s_W\sqrt{3-t_W^2}}$
\\
\hline
$Z \phi^\pm \phi^\mp$ & $\pm \frac{c_W^2-s_W^2}{2s_Wc_W} \mp \delta_Z\frac{1-t_W^2}{2s_W\sqrt{3-t_W^2}}$
\\
\hline
$Z' x^\pm x^\mp$ & $\mp \frac{1-t_W^2}{2s_W\sqrt{3-t_W^2}} \mp \delta_Z \frac{c_W^2-s_W^2}{2s_Wc_W}$
\\
\hline
$Z' \phi^\pm \phi^\mp$ & $\mp \frac{1-t_W^2}{2s_W\sqrt{3-t_W^2}} \mp \delta_Z \frac{c_W^2-s_W^2}{2s_Wc_W}$
\\
\hline
\end{tabular}
\end{center}
\caption{Vertices $[{\rm SV_\mu V_\nu}]=\ii e K g^{\mu\nu}$ and $[V_\mu S(p_1)S(p_2)]=\ii e G(p_1-p_2)^\mu$.\label{tab:SVV.VSS}}
\end{table}

\subsubsection{Vector Boson Lagrangian}

From the Lagrangian
\begin{equation}
\mathcal{L}_V=-\frac{1}{2}\textrm{Tr}\{\widetilde{G}_{\mu\nu}\widetilde{G}^{\mu\nu}\}-\frac{1}{4}B_x^{\mu\nu}B_{x \, \mu\nu}\,, \qquad \widetilde{G}_{\mu\nu}=\frac{\ii}{g}[D_\mu,D_\nu]\, ,
\end{equation}
and using (\ref{vecmix}) and (\ref{zmix}) we obtain, in a straightforward way, the [VVV] couplings we need for our study of LFV processes. They are given in table~\ref{tab:VVV}.

\begin{table}
\begin{center}
\begin{tabular}{|c|c||c|c|}
\hline
VVV & J & VVV & J
\\
\hline
$\gamma X^+ X^-$ & $-1$ & $\gamma W^+W^-$ & $-1$
\\
\hline
$Z X^+ X^- $ & $\frac{c_W^2-s_W^2}{2s_Wc_W}-\frac{\delta_Z}{2s_W}\sqrt{3-t_W^2}$ & $ZW^+W^-$ & $\frac{c_W}{s_W}$
\\
\hline
$Z' X^+ X^-$ & $-\frac{1}{2s_W}\sqrt{3-t_W^2}-\delta_Z \frac{c_W^2-s_W^2}{2s_Wc_W}$ & $Z' W^+W^-$ & $-\delta_Z\frac{c_W}{s_W}$
\\
\hline
\end{tabular}
\end{center}
\caption{Vertices $[{\rm V}_\mu(p_1) {\rm V}_\nu(p_2) {\rm V}_\rho(p_3)]=\ii J [g^{\mu\nu}(p_2-p_1)^\rho+g^{\nu\rho}(p_3-p_2)^\mu+g^{\mu\rho}(p_1-p_3)^\nu]$.\label{tab:VVV}}
\end{table}

\subsection{Lepton sector}

Here we will treat the pieces involving leptons which include their Yukawa Lagrangian as well as their gauge interactions.  We will obtain the lepton masses and physical states and their couplings to gauge and Goldstone bosons.

\subsubsection{Lepton Yukawa sector}

Lepton masses follow from the Yukawa Lagrangian:
\begin{equation}
\mathcal{L}_Y\supset \ii\lambda_N^m \bar{N}_{Rm} \Phi_2^\dagger L_m + \frac{\ii\lambda_\ell^{mn}}{\Lambda} \bar{\ell}_{Rm} \epsilon_{ijk} \Phi_1^i \Phi_2^j L_n^k + \textrm{h.c.}\, ,
\end{equation}
where the quartic term preserves the global symmetry ($L_m$ transforms as ({\bf 1},{\bf 3}) under $SU(3)_1\times SU(3)_2$) and $\lambda_N$ can be taken diagonal after a proper field redefinition.
Firstly we need to determine the actual physical states of the leptons.   Keeping only the mass terms to order $\mathcal{O}(v^2/f^2)$ we have the following Lagrangian:
\begin{equation}
\begin{split}
\mathcal{L}_Y & \supset -fs_\beta\lambda_N^m\left[\left(1-\frac{\delta_\nu^2}{2}\right)\bar{N}_{Rm}N_{Lm}-
\delta_\nu\bar{N}_{Rm}\nu_{Lm}\right]+\zeta_\beta\frac{fv}{\sqrt{2}\Lambda}\lambda_\ell^{mn}\bar{\ell}_{Rm}\ell_{Ln} + \textrm{h.c.}\, ,
\label{lepmix1}
\end{split}
\end{equation}
where
\bea
\delta_\nu=-\frac{v}{\sqrt{2}ft_\beta}\ ,\quad
\zeta_\beta=\left[1-\frac{v^2}{4f^2}-\frac{v^2}{12f^2}\left(\frac{s_\beta^4}{c_\beta^2}+\frac{c_\beta^4}{s_\beta^2}\right)\right]\ .
\eea

The matrices $\lambda_N$ and $\lambda_\ell$ are not necessarily aligned. 
Thus, in the basis where the former is diagonal, the latter mixes different light lepton flavors. 
Denoting the eigenvalues of $\lambda_\ell$ as $y_{\ell_i}$, the light lepton masses are given by
\bea
m_{\ell_i}=-\zeta_\beta\frac{fv}{\sqrt{2}\Lambda}y_{\ell_i}\ ,
\eea
whereas the left-handed components of the light physical fields are obtained by the replacement:
\bea
\ell_{Lm} \to (V_\ell \ell_L)_m = V^{mi}_\ell\ell_{Li}\ .
\label{lep1}
\eea
Furthermore, according to (\ref{lepmix1}) each heavy neutrino is mixed just with the light neutrino of the same family.  To separate them we rotate only the left-handed sector.  To order $\mathcal{O}(v^2/f^2)$, the physical states for the neutrinos are given by:
\begin{equation}
\left(\begin{array}{c} \nu_L \\ N_L \end{array}\right)_m  \to 
\left[\left(\begin{array}{cc} 1-\frac{\delta_\nu^2}{2} & -\delta_\nu \\
\delta_\nu & 1-	\frac{\delta_\nu^2}{2} \end{array}\right)
\left(\begin{array}{c} V_\ell\nu_L \\ N_L \end{array}\right)\right]_m\, .
\label{lep2}
\end{equation}
Notice that the mixing angle $\delta_\nu$ between light doublet and heavy singlet neutrinos is experimentally constrained to be small \cite{delAguila:2008pw,delAguila:2009bb}. Although the bound is flavor dependent (0.05, 0.03 and 0.09 at 95\% C.L. for $\nu_e$, $\nu_\mu$ and $\nu_\tau$ respectively) we will assume a typical upper limit $\delta_\nu<0.05$ for illustration purposes.

Since one can safely consider the SM neutrinos as massless, we have chosen to rotate them in the same way as the light charged leptons. 
Finally, the heavy neutrino masses are:
\begin{equation}
m_{N_i}=fs_\beta\lambda_N^i\, .
\end{equation}

We now need the Goldstone-lepton couplings.  These require expansions of up to order $\mathcal{O}(v^4/f^4)$ in order to get the couplings to order $\mathcal{O}(v^2/f^2)$.  Goldstone rotations from the scalar sector in (\ref{goldrot}) are also needed.  We want only terms that couple one charged Goldstone boson ($x^\pm$ and $\phi^\pm$) with two fermions.  Taking these considerations into account and using the physical fermion states (\ref{lep1}) and (\ref{lep2}) we obtain the couplings in table \ref{tab:SFF}. Note that several coupligs are zero because we have neglected the SM neutrino masses.

\begin{table}
\begin{center}
\begin{tabular}{|c|c|c|}
\hline
SFF & $c_L$ & $c_R$
\\
\hline
$x^+ \bar{N}_i \ell_j$ & $-\frac{1}{\sqrt{2}s_W}\frac{m_{N_i}}{M_X}\left(1-\frac{\delta_\nu^2}{2}\right)V_{\ell}^{ij}$ & $\frac{1}{\sqrt{2}s_W}\frac{m_{\ell_j}}{M_X}\left(1-\delta_\nu^2\right)V_{\ell}^{ij}$ 
\\
\hline
$x^- \bar{\ell}_j N_i$ & $\frac{1}{\sqrt{2}s_W}\frac{m_{\ell_j}}{M_X}\left(1-\delta_\nu^2\right)V_{\ell}^{ij*}$ & $-\frac{1}{\sqrt{2}s_W}\frac{{m_N}_i}{M_X}\left(1-\frac{\delta_\nu^2}{2}\right)V_{\ell}^{ij*}$
\\
\hline
$\phi^+ \bar{N}_i \ell_j$ & $\delta_\nu \frac{\ii}{\sqrt{2}s_W}\frac{{m_N}_i}{M_W} V_{\ell}^{ij}$ & $\delta_\nu\frac{\ii}{\sqrt{2}s_W}\frac{m_{\ell_j}}{M_W}V_{\ell}^{ij}$
\\
\hline
$\phi^- \bar{\ell}_j N_i$ & $-\delta_\nu\frac{\ii}{\sqrt{2}s_W}\frac{m_{\ell_j}}{M_W}V_{\ell}^{ij*}$ &
$-\delta_\nu \frac{\ii}{\sqrt{2}s_W}\frac{{m_N}_i}{M_W} V_{\ell}^{ij*}$
\\
\hline
$x^+ \bar{\nu}_i \ell_i$ & $0$ & $-\delta_\nu \frac{1}{\sqrt{2}s_W} \frac{m_{\ell_i}}{M_X}$
\\
\hline
$x^-\bar{\ell}_i \nu_i$ & $-\delta_\nu \frac{1}{\sqrt{2}s_W} \frac{m_{\ell_i}}{M_X}$ & $0$
\\
\hline
$\phi^+ \bar{\nu}_i \ell_i$ & $0$ & $\frac{\ii}{\sqrt{2}s_W}\frac{m_{\ell_i}}{M_W} \left( 1-\frac{\delta_\nu^2}{2} \right)$
\\
\hline
$\phi^- \bar{\ell}_i \nu_i$ & $-\frac{\ii}{\sqrt{2}s_W}\frac{m_{\ell_i}}{M_W} \left( 1-\frac{\delta_\nu^2}{2} \right)$ & $0$
\\
\hline
\end{tabular}
\end{center}
\caption{Vertices ${\rm [SFF]}=\ii e (c_L P_L+c_RP_R)$ for the lepton sector.\label{tab:SFF}}
\end{table}

\subsubsection{Lepton-Gauge sector}

\begin{equation}
\mathcal{L}_F = \bar{\psi}_m\ii\cancel{D}\psi_m\ , \qquad \psi_m=\{ L_m, {\ell_{Rm}}, {N_{Rm}} \}\ .
\end{equation}
The expression for the covariant derivative is in equation (\ref{covar}).  We note that the only places where $v^2/f^2$ corrections appear are in the definitions of the physical states of the leptons (in $\delta_\nu^2$) and in the $ZZ'$ mixing (in $\delta_Z$).
We must use (\ref{lep1}) and (\ref{lep2}) and the definitions of the physical gauge bosons to obtain the relevant Feynman rules which are listed in table~\ref{tab:VFF}.

\begin{table}
\begin{center}
\begin{tabular}{|c|c|c|}
\hline
VFF & $g_L$ & $g_R$ 
\\
\hline
$\gamma \bar{\ell}_i \ell_i$ & 1 & 1 
\\
\hline
$W^+\bar{\nu}_i \ell_i$ & $\frac{1}{\sqrt{2}s_W}\left(1-\frac{\delta_\nu^2}{2}\right)$ & 0
\\
\hline
$W^+ \bar{N}_m \ell_i$ &
$-\delta_\nu\frac{1}{\sqrt{2}s_W}V_{\ell}^{mi}$ & 0
\\
\hline
$Z\bar{\ell}_i \ell_i$ & $\frac{-1+2s_W^2}{2s_Wc_W} + \delta_Z \frac{1-2s_W^2}{2s_Wc_W^2\sqrt{3-t_W^2}}$ & $\frac{s_W}{c_W} - \delta_Z \frac{s_W}{c_W^2\sqrt{3-t_W^2}}$ 
\\
\hline
$Z \bar{\nu}_i \nu_i$ & $\frac{1}{2s_Wc_W}(1-\delta_\nu^2)+ \delta_Z \frac{1-2s_W^2}{2s_Wc_W^2\sqrt{3-t_W^2}}$ & 0 
\\
\hline
$Z \bar{N}_i N_i$ & $\delta_\nu^2\frac{1}{2s_Wc_W} - \delta_Z \frac{1}{s_W\sqrt{3-t_W^2}}$ & 0 
\\
\hline
$Z \bar{N}_m \nu_i$ & $-\delta_\nu\frac{1}{2s_Wc_W} V_{\ell}^{mi}$ & 0 
\\
\hline
$X^+ \bar{\nu}_i \ell_i$ & $-\delta_\nu\frac{\ii}{\sqrt{2}s_W} $ & 0
\\
\hline
$X^+ \bar{N}_m \ell_i$ & $-\frac{\ii}{\sqrt{2}s_W}\left(1-\frac{\delta_\nu^2}{2}\right)V_{\ell}^{mi}$ & 0
\\
\hline
$Y^0\bar{\nu}_i \nu_i$ & $\delta_\nu\frac{\ii}{\sqrt{2}s_W}$ & 0
\\
\hline
$Y^0 \bar{N}_i N_i$ & $-\delta_\nu\frac{\ii}{\sqrt{2}s_W}$ & 0
\\
\hline
$Y^0 \bar{\nu}_i N_m$ & $\frac{\ii}{\sqrt{2}s_W}(1-\delta_\nu^2)V_{\ell}^{mi*}$ & 0
\\
\hline
$Y^0 \bar{N}_m \nu_i$ & $-\delta_\nu^2 \frac{\ii}{\sqrt{2}s_W}V_{\ell}^{mi}$ & 0
\\
\hline
$Z'\bar{\ell}_i \ell_i$ & $\frac{1-2s_W^2}{2s_Wc_W^2\sqrt{3-t_W^2}}-\delta_Z \frac{-1+2s_W^2}{2s_Wc_W}$ & $-\frac{s_W}{c_W^2\sqrt{3-t_W^2}} - \delta_Z \frac{s_W}{c_W}$
\\
\hline
$Z' \bar{\nu}_i \nu_i$ & $\frac{1-2s_W^2}{2s_Wc_W^2\sqrt{3-t_W^2}} \left(1-\delta_\nu^2\frac{c_W^2(3-t_W^2)}{1-2s_W^2}\right) - \delta_Z \frac{1}{2s_Wc_W}$ & 0
\\
\hline
$Z' \bar{N}_i N_i$  & $-\frac{1}{s_W\sqrt{3-t_W^2}}\left(1-\delta_\nu^2\frac{3-t_W^2}{2}\right)$ & 0
\\
\hline
$Z' \bar{N}_m \nu_i$ & $-\delta_\nu \frac{1}{2s_W}\sqrt{3-t_W^2}V_{\ell}^{mi}$ & 0
\\
\hline
\end{tabular}
\end{center}
\caption{Vertices ${\rm [VFF]}=\ii e (g_LP_L+g_RP_R)$ for the lepton sector. Notice that $Y^0\ne Y^{0\dagger}$. \label{tab:VFF}}
\end{table}

\subsection{Quark sector}

For the $\mueN$ conversion process we also require some of the quark couplings.  The full mixing structure of the quark sector is much more complex than that of the lepton sector and, in general, all light quarks mix with other heavy and light quarks from every family.  However, we are only interested in the mixing effects that are a consequence of mixing in the lepton sector so we will neglect most mixing effects of the quark sector.  Corrections of order $v^2/f^2$ to vertices are only needed for particles involved in triangle diagrams and, since quarks only appear in box diagrams, $v/f$ precision is sufficient.  This simplifies the calculation of our Feynman rules considerably.  However, we do have to analyze both the universal and anomaly-free embeddings since they produce different results in general.

\subsubsection{Quark Yukawa sector}

For the \emph{anomaly-free} embedding, the basic Yukawa Lagrangian reads:
\begin{eqnarray}
\mathcal{L}_Y& \supset & \lambda_1^t \bar{u}_{R3}^1 \Phi_1^\dag Q_3 + \ii\lambda_2^t \bar{u}_{R3}^2 \Phi_2^\dag Q_3 + \ii\frac{\lambda_b^m}{\Lambda}\bar{d}_{Rm}\epsilon_{ijk}
\Phi_1^i\Phi_2^jQ_3^k\nn \\
 && + \ii\lambda_1^{dn} \bar{d}_{Rn}^1 Q_n^T \Phi_1 + \ii\lambda_2^{dn} \bar{d}_{Rn}^2 Q_n^T \Phi_2 + \ii\frac{\lambda_u^{mn}}{\Lambda}\bar{u}_{Rm}\epsilon_{ijk}
\Phi_1^{*i}\Phi_2^{*j}Q_n^k\,,
\end{eqnarray}
where $n=1,2$; $i,j,k=1,2,3$ are $SU(3)$ indices; $d_{Rm}$ runs over $(d_R,s_R,b_R,D_R,S_R)$ and $u_{Rm}$ runs over $(u_R,c_R,t_R,T_R)$;  $u_{R3}^1$ and $u_{R3}^2$ are linear combinations of $t_R$ and $T_R$; 
$d_{R1}^n$ and $d_{R2}^n$ are linear combinations of $d_R$ and $D_R$ for $n=1$ and of $s_R$ and $S_R$ for $n=2$:
\begin{align}
{T}_R& = \frac{\lambda_1^tc_\beta {u}_{R3}^1 + \lambda^t_2 s_\beta {u}_{R3}^2}{
\sqrt{{(\lambda_1^t)}^2c_\beta^2 + {(\lambda_2^t)}^2s_\beta^2}}\ ,& 
{t}_R &=  \frac{-\lambda_2^ts_\beta {u}_{R3}^1 + \lambda^t_1 c_\beta {u}_{R3}^2}{
\sqrt{{(\lambda_1^t)}^2c_\beta^2 + {(\lambda_2^t)}^2s_\beta^2}}\ , \\
{D}_R& =  \frac{\lambda_1^{d1}c_\beta {d}_{R1}^1 + \lambda^{d1}_2 s_\beta {d}_{R1}^2}{
\sqrt{{(\lambda_1^{d1})}^2c_\beta^2 + {(\lambda_2^{d1})}^2s_\beta^2}}\ ,& 
{d}_R &=  \frac{-\lambda_2^{d1}s_\beta {d}_{R1}^1 + \lambda^{d1}_1 c_\beta {d}_{R1}^2}{
\sqrt{{(\lambda_1^{d1})}^2c_\beta^2 + {(\lambda_2^{d1})}^2s_\beta^2}}\ , \\
{S}_R &=  \frac{\lambda_1^{d2}c_\beta {d}_{R2}^1 + \lambda^{d2}_2 s_\beta {d}_{R2}^2}{
\sqrt{{(\lambda_1^{d2})}^2c_\beta^2 + {(\lambda_2^{d2})}^2s_\beta^2}}\ ,& 
{s}_R &=  \frac{-\lambda_2^{d2}s_\beta {d}_{R2}^1 + \lambda^{d2}_1 c_\beta {d}_{R2}^2}{
\sqrt{{(\lambda_1^{d2})}^2c_\beta^2 + {(\lambda_2^{d2})}^2s_\beta^2}}\, .
\end{align}
We require the collective structure with different right-handed quarks 
entering in the $\Phi_1$ and $\Phi_2$ quartic Yukawa couplings. 
By a proper field redefinition, $\lambda_1^{d}$ can be taken diagonal in general and, for simplicity and to avoid large quark flavor changing effects, we also 
assume $\lambda_2^{d}$ to be diagonal \cite{Kaplan:2003uc,Han:2005ru}.

Before the EWSB we obtain the following masses for the heavy quarks:
\begin{eqnarray}
m_T & = & f\sqrt{{(\lambda_1^t)}^2c_\beta^2 + {(\lambda_2^t)}^2s_\beta^2}\ , \\
m_D & = & f\sqrt{{(\lambda_1^{d1})}^2c_\beta^2 + {(\lambda_2^{d1})}^2s_\beta^2}\ , \\
m_S & = & f\sqrt{{(\lambda_1^{d2})}^2c_\beta^2 + {(\lambda_2^{d2})}^2s_\beta^2}\ .
\end{eqnarray}

After the EWSB, the quark mass terms work out as follows to leading order:
\begin{eqnarray}
\mathcal{L}_Y & \supset & -m_T \bar{T}_R T_L 
+ \frac{v}{\sqrt{2}} \frac{s_\beta c_\beta [{(\lambda^t_1)}^2 - {(\lambda^t_2)}^2]}{
\sqrt{{(\lambda_1^t)}^2c_\beta^2 + {(\lambda_2^t)}^2s_\beta^2}} \bar{T}_R t_L 
- \frac{v}{\sqrt{2}} \frac{\lambda_1^t\lambda_2^t}{
\sqrt{{(\lambda_1^t)}^2c_\beta^2 + {(\lambda_2^t)}^2s_\beta^2}} \bar{t}_R t_L \nonumber \\
&& + \frac{v}{\sqrt{2}}\frac{f}{\Lambda} \lambda_u^{mn} \bar{u}_{Rm} u_{Ln}  \nonumber \\
&& -m_D \bar{D}_R D_L - \frac{v}{\sqrt{2}} \frac{s_\beta c_\beta ({(\lambda^{d1}_1)}^2-{(\lambda^{d1}_2)}^2)}{
\sqrt{{(\lambda^{d1}_1)}^2 c_\beta^2 + {(\lambda^{d1}_2)}^2 s_\beta^2}} \bar{D}_R d_L + \frac{v}{\sqrt{2}}\frac{\lambda^{d1}_1\lambda^{d1}_2}{
\sqrt{{(\lambda^{d1}_1)}^2 c_\beta^2 + {(\lambda^{d1}_2)}^2 s_\beta^2}} \bar{d}_R d_L \nonumber \\
&& -m_S \bar{S}_R S_L - \frac{v}{\sqrt{2}} \frac{s_\beta c_\beta ({(\lambda^{d2}_1)}^2-{(\lambda^{d2}_2)}^2)}{
\sqrt{{(\lambda^{d2}_1)}^2 c_\beta^2 + {(\lambda^{d2}_2)}^2 s_\beta^2}} \bar{S}_R s_L + \frac{v}{\sqrt{2}}\frac{\lambda^{d2}_1\lambda^{d2}_2}{
\sqrt{{(\lambda^{d2}_1)}^2 c_\beta^2 + {(\lambda^{d2}_2)}^2 s_\beta^2}} \bar{s}_R s_L \nonumber \\
&& +\frac{v}{\sqrt{2}}\frac{f}{\Lambda} \lambda^m_b \bar{d}_{Rm} b_L + \textrm{h.c.}\, .
\end{eqnarray}

Since we are interested in lepton flavor mixing we will assume no flavor mixing in the quark sector, which might otherwise dilute some of the effects we wish to highlight.  We essentially set all the $\lambda_b^m$ and $\lambda_u^{mn}$ that mix different families or heavy and light quarks to zero and all others are fixed by the light quark masses.  Only the heavy-light mixing within each family remains.  We neglect terms proportional to $v^2/f^2$ and rotate the left-handed fields to obtain the physical quark states (heavy quark masses get corrections at this order that will be neglected as well):
\begin{eqnarray}
T_L & \to & T_L + \delta_t t_L\ , \\
t_L & \to & t_L - \delta_t T_L\ , \\
D_{L} & \to & D_{L} + \delta_{d} d_{L}\ , \\
d_{L} & \to & d_{L} -\delta_{d} D_{L}\ , \\
S_{L} & \to & S_{L} + \delta_{s} s_{L}\ , \\
s_{L} & \to & s_{L} -\delta_{s} S_{L}\ ,
\end{eqnarray}
where
\begin{eqnarray}
\delta_t & = & \frac{v}{\sqrt{2}f} \frac{s_\beta c_\beta [{(\lambda_1^{t})}^2-{(\lambda_2^{t})}^2]}{\sqrt{{(\lambda_1^t)}^2c_\beta^2-{(\lambda_2^t)}^2s_\beta^2}}\ , \\
\delta_d & = & -\frac{v}{\sqrt{2}f} \frac{s_\beta c_\beta [{(\lambda_1^{d1})}^2-{(\lambda_2^{d1})}^2]}{\sqrt{{(\lambda_1^{d1})}^2c_\beta^2-{(\lambda_2^{d1})}^2s_\beta^2}}\ , \\
\delta_s & = & -\frac{v}{\sqrt{2}f} \frac{s_\beta c_\beta [{(\lambda_1^{d2})}^2-{(\lambda_2^{d2})}^2]}{\sqrt{{(\lambda_1^{d2})}^2c_\beta^2-{(\lambda_2^{d2})}^2s_\beta^2}}
\end{eqnarray}
are complex in general.

Taking all this into account we get the SM quark masses:
\begin{eqnarray}
m_u \,=\, -\frac{v}{\sqrt{2}}\frac{f}{\Lambda}\lambda_u^{11}\ , &&
m_c \,=\, -\frac{v}{\sqrt{2}}\frac{f}{\Lambda}\lambda_u^{22}\ , \qquad
m_b \,=\, -\frac{v}{\sqrt{2}}\frac{f}{\Lambda}\lambda_b^{3}\ , \\
m_t & = & \frac{v}{\sqrt{2}}\frac{\lambda_1^t\lambda_2^t}{\sqrt{{(\lambda_1^t)}^2c_\beta^2 + {(\lambda_2^t)}^2s_\beta^2}}\ , \\
m_d & = & -\frac{v}{\sqrt{2}}\frac{\lambda^{d1}_1\lambda^{d1}_2}{\sqrt{{(\lambda^{d1}_1)}^2 c_\beta^2 
+ {(\lambda^{d1}_2)}^2 s_\beta^2}}\ , \\
m_s & = & -\frac{v}{\sqrt{2}}\frac{\lambda^{d2}_1\lambda^{d2}_2}{
\sqrt{{(\lambda^{d2}_1)}^2 c_\beta^2 + {(\lambda^{d2}_2)}^2 s_\beta^2}}\ .
\end{eqnarray}

The $\delta_{d,s,t}$ parameters can be expressed in terms of the quark masses:
\begin{eqnarray}
 \delta_q & = & \pm \frac{vm_Q}{2\sqrt{2}f^2}\frac{1}{c_\beta s_\beta} \left( s_\beta^2 - c_\beta^2 + \epsilon 
 \sqrt{1-\frac{8c_\beta^2 s_\beta^2 f^2 m_q^2}{v^2m_Q^2}} \right) \ , 
\label{deltaq}
\end{eqnarray}
where the $+\ (-)$ sign stands for $q = t\ (d,s)$ and $\epsilon = \pm1$ depending on the corresponding 
values of $\lambda_1$ and $\lambda_2$. We will use this expression to study the predictions of the 
model because it involves parameters with a more straightforward physical meaning. 

With this information and redefining the Goldstone fields as given in (\ref{goldrot}), we can obtain the relevant
quark-Goldstone boson couplings for our processes. Those involving first family quarks are given in table~\ref{tab:SFF-af}. Remember that we have removed all quark flavor changing vertices so there is no CKM matrix.

\begin{table}
\begin{center}
\begin{tabular}{|c|c|c|}
\hline
SFF & $c_L$ & $c_R$
\\
\hline
$x^-\bar{D}u$ & $-\frac{1}{\sqrt{2}s_W} \frac{m_D}{M_X}$ & $\frac{1}{\sqrt{2}s_W} \frac{m_u}{M_X}$
\\
\hline
$x^-\bar{d}u$ & 0 & $\delta_d^*\frac{1}{\sqrt{2}s_W} \frac{m_d}{M_X}$
\\
\hline
\end{tabular}
\begin{tabular}{|c|c|c|}
\hline
SFF & $c_L$ & $c_R$
\\
\hline
$\phi^-\bar{D}u$ & $\delta_d\frac{\ii}{\sqrt{2}s_W} \frac{m_D}{M_W}$ & $-\delta_d^*\frac{\ii}{\sqrt{2}s_W} \frac{m_u}{M_W}$
\\
\hline
$\phi^-\bar{d}u$ & $-\frac{\ii}{\sqrt{2}s_W}\frac{m_d}{M_W}$  & $\frac{\ii}{\sqrt{2}s_W} \frac{m_u}{M_W}$ 
\\
\hline
\end{tabular}
\end{center}
\caption{Vertices ${\rm [SFF]}=\ii e (c_LP_L+c_RP_R)$ involving first family quarks in the anomaly-free embedding.\label{tab:SFF-af}}
\end{table}

The situation is similar in the \emph{universal} embedding although the Yukawa Lagrangian is different:
\begin{equation}
\mathcal{L}_Y \supset  \ii \lambda^{un}_1 \bar{u}_{Rn}^1 \Phi_1^\dag Q_n + \ii \lambda^{un}_2 \bar{u}_{Rn}^2 \Phi_2^\dag Q_n
+\ii\frac{\lambda_d^{mn}}{\Lambda} \bar{d}_{Rm} \epsilon_{ijk} \Phi_1^i \Phi_2^j Q_n^k + {\rm h.c.}\ .
\end{equation}
Here $m,n=1,2,3$ are generation indices and $i,j,k=1,2,3$ are $SU(3)$ indices;  $d_m$ runs over the down quarks $(d,s,b)$ and $u^{1,2}_n$ are linear combinations of the light and heavy up quarks:
\begin{eqnarray}
U_{Rn} & = & \frac{\lambda_1^{un}c_\beta u_{Rn}^1 + \lambda_2^{un}s_\beta u_{Rn}^2}{\sqrt{{(\lambda_1^{un})}^2c_\beta^2 + {(\lambda_2^{un})}^2s_\beta^2}}\ ,\\
u_{Rn} & = & \frac{-\lambda_1^{un}s_\beta u_{Rn}^1 + \lambda_2^{un}c_\beta u_{Rn}^2}{\sqrt{{(\lambda_1^{un})}^2c_\beta^2 + {(\lambda_2^{un})}^2s_\beta^2}}\ .
\end{eqnarray}
Analogously to the anomaly free case, we assume the collective structure. As before, $\lambda^u_1$ can be made diagonal by a field redefinition and $\lambda^u_2$ is also taken diagonal to 
avoid large quark flavor effects.

The mass terms work out as:
\begin{eqnarray}
\mathcal{L}_Y & \supset & -f\sqrt{{(\lambda_1^{un})}^2 c_\beta^2 + {(\lambda_2^{un})}^2 s_\beta^2} \ \bar{U}_{Rn} U_{Ln} +
\frac{v}{\sqrt{2}}\frac{s_\beta c_\beta \left[ {(\lambda_1^{un})}^2-{(\lambda_2^{un})}^2 \right]}{\sqrt{{(\lambda_1^{un})}^2 c_\beta^2 + {(\lambda_2^{un})}^2 s_\beta^2}} \ \bar{U}_{Rn} u_{Ln} \nonumber \\
&& - \frac{v}{\sqrt{2}}\frac{\lambda_1^{un}\lambda_2^{un}}{\sqrt{{(\lambda_1^{un})}^2 c_\beta^2 + {(\lambda_2^{un})}^2 s_\beta^2}} \ \bar{u}_{Rn} u_{Ln} 
+ \frac{vf}{\sqrt{2}\Lambda} \lambda_d^{ij} \bar{d}_{Ri}d_{Lj} + \textrm{h.c.}\ .
\end{eqnarray}
We have neglected terms proportional to $v^2/f$.

We will again ignore all generation mixing terms.  This means setting $\lambda_d^{ij} = \lambda_d^i \delta_{ij}$.  The only remaining mixing terms involve the light and heavy up quarks of each generation.  The following rotation of the left-handed fields is required to obtain diagonal mass terms:
\begin{eqnarray}
U_{Ln} & \to & U_{Ln} + \delta_{u_n} u_{Ln} \ , \\
u_{Ln} & \to & u_{Ln} - \delta_{u_n} U_{Ln} \ ,
\end{eqnarray}
where
\begin{equation}
\delta_{u_n} = \frac{v}{\sqrt{2}f}\frac{s_\beta c_\beta [{(\lambda_1^{un})}^2-{(\lambda_2^{un})}^2]}{(\lambda_1^{un})^2 c_\beta^2 + (\lambda_2^{un})^2 s_\beta^2} \ .
\end{equation}
The quark masses to order $v/f$ are:
\begin{eqnarray}
m_{U_n} & = & f \sqrt{(\lambda_1^{un})^2 c_\beta^2 + (\lambda_2^{un})^2 s_\beta^2}\ , \\
m_{u_n} & = & \frac{v}{\sqrt{2}}\frac{\lambda_1^{un}\lambda_2^{un}}{\sqrt{(\lambda_1^{un})^2 c_\beta^2 + (\lambda_2^{un})^2 s_\beta^2}}\ , \\
m_{d_n} & = & \frac{vf}{\sqrt{2}\Lambda} \lambda_d^{n}\ .
\end{eqnarray}
Here $\delta_{u_n}$ can also be written analogously to (\ref{deltaq}) using the global + sign and with $\epsilon=\pm 1$ depending on the values of $\lambda_{1,2}$.

The Feynman rules for quark-Goldstone couplings to order $v/f$ are given in table~\ref{tab:SFF-u}.

\begin{table}
\begin{center}
\begin{tabular}{|c|c|c|}
\hline
SFF & $c_L$ & $c_R$
\\
\hline
$x^+ \bar{u}_id_i$ & 0 & $\delta_{u_i}^* \frac{1}{\sqrt{2}s_W} \frac{m_{d_i}}{M_X}$
\\
\hline
$x^+\bar{U}_id_i$ & $-\frac{1}{\sqrt{2}s_W} \frac{m_{U_i}}{M_X}$ & $\frac{1}{\sqrt{2}s_W} \frac{m_{d_i}}{M_X}$
\\
\hline
\end{tabular}
\begin{tabular}{|c|c|c|}
\hline
SFF & $c_L$ & $c_R$
\\
\hline
$\phi^+\bar{u}_id_i$ & $-\frac{\ii}{\sqrt{2}s_W} \frac{m_{u_i}}{M_W}$ & $\frac{\ii}{\sqrt{2}s_W} \frac{m_{d_i}}{M_W}$
\\
\hline
$\phi^+ \bar{U}_id_i$ & $\ii\delta_{u_i}\frac{1}{\sqrt{2}s_W}\frac{m_{U_i}}{M_W}$ & $-\ii\delta_{u_i}^* \frac{1}{\sqrt{2}s_W} \frac{m_{d_i}}{M_W}$ 
\\
\hline
\end{tabular}
\end{center}
\caption{Vertices ${\rm [SFF]}=\ii e (c_LP_L+c_RP_R)$ involving first family quarks in the universal embedding.\label{tab:SFF-u}}
\end{table}

\subsubsection{Quark-Gauge sector}

For the anomaly-free embedding the Lagrangian is:
\begin{equation}
\mathcal{L}_F = \bar{Q}_m\ii\cancel{D}_m^L Q_m + \bar{u}_{Rm}\ii\cancel{D}^u u_{Rm} + \bar{d}_{Rm}\ii\cancel{D}^d d_{Rm}
+ \bar{T}_{R}\ii\cancel{D}^u T_{R} + \bar{D}_{R}\ii\cancel{D}^d D_{R} + \bar{S}_{R}\ii\cancel{D}^d S_{R}\ .
\end{equation}
Here we have taken into account that the first two families are in the anti-fundamental representation:
\begin{eqnarray}
D_{\{1,2\}\mu}^L & = & \partial_\mu + \ii gA_\mu^aT_a^*\ , \\
D_{3\mu}^L & = & \partial_\mu - \ii gA_\mu^aT_a + ig_x\frac{1}{3}B^x_\mu\ , \\
D_{\mu}^u & = & \partial_\mu + \ii g_x \left(-\frac{1}{3}\right) B^x_\mu\ , \\
D_{\mu}^d & = & \partial_\mu + \ii g_x \frac{2}{3} B^x_\mu\ .
\end{eqnarray}

For the universal embedding, the Lagrangian is more symmetric:
\begin{equation}
\mathcal{L} = \bar{Q}_m\ii\cancel{D}^L Q_m + \bar{u}_{Rm}\ii\cancel{D}^u u_{Rm} + \bar{d}_{Rm}\ii\cancel{D}^d d_{Rm} + \bar{U}_{Rm}\ii\cancel{D}^u U_{Rm}\ ,
\end{equation}
where
\begin{eqnarray}
D_{\mu}^L & = & \partial_\mu - \ii gA_\mu^aT_a + ig_x\frac{1}{3}B^x_\mu\ , \\
D_{\mu}^u & = & \partial_\mu + \ii g_x \left(-\frac{1}{3}\right) B^x_\mu\ , \\
D_{\mu}^d & = & \partial_\mu + \ii g_x \frac{2}{3} B^x_\mu\ .
\end{eqnarray}

The Feynman rules for the gauge-fermion couplings in both embeddings are given in tables~\ref{tab:VFF-af} and \ref{tab:VFF-u}. We only include the vertices and order relevant for our calculations.

\begin{table}
\begin{center}
\begin{tabular}{|c|c|c|}
\hline
VFF & $g_L$ & $g_R$ 
\\
\hline
$\gamma \bar{u} u$ & $-\frac{2}{3}$ & $-\frac{2}{3}$ 
\\
\hline
$\gamma \bar{d} d$ & $\frac{1}{3}$ & $\frac{1}{3}$ 
\\
\hline
$W^-\bar{D} u$ & $-\delta_d^* \frac{1}{\sqrt{2}s_W} $ & 0
\\
\hline
$W^-\bar{d} u$ & $\frac{1}{\sqrt{2}s_W}$ & 0
\\
\hline
$Z \bar{u} u$ & $\frac{-1+4c_W^2}{6c_Ws_W}$ & $\frac{-2s_W}{3c_W}$ 
\\
\hline
$Z \bar{d} d$ & $-\frac{1+2c_W^2}{6c_Ws_W}$ & $\frac{s_W}{3c_W}$ 
\\
\hline
$X^- \bar{D} u$ & $-\frac{\ii}{\sqrt{2}s_W}$ & 0
\\
\hline
$X^- \bar{d} u$ & $-\delta_d^* \frac{1}{\sqrt{2}s_W}$ & 0
\\
\hline
$Z' \bar{u} u$ & $-\frac{\sqrt{3-t_W^2}}{6s_W}$ & $\frac{2t_W^2}{3s_W\sqrt{3-t_W^2}}$ 
\\
\hline
$Z' \bar{d} d$ & $-\frac{\sqrt{3-t_W^2}}{6s_W}$ & $-\frac{t_W^2}{3s_W\sqrt{3-t_W^2}}$  
\\
\hline
\end{tabular}
\end{center}
\caption{Vertices ${\rm [VFF]}=\ii e (g_LP_L+g_RP_R)$ for the quark sector in the anomaly-free embedding 
entering in our calculation. \label{tab:VFF-af}}
\end{table}

\begin{table}
\begin{center}
\begin{tabular}{|c|c|c|}
\hline
VFF & $g_L$ & $g_R$ 
\\
\hline
$\gamma \bar{u} u$ & $-\frac{2}{3}$ & $-\frac{2}{3}$ 
\\
\hline
$\gamma \bar{d} d$ & $\frac{1}{3}$ & $\frac{1}{3}$ 
\\
\hline
$W^+\bar{U} d$ & $-\delta_{u_i}^* \frac{1}{\sqrt{2}s_W}$ & 0
\\
\hline
$W^+\bar{u} d$ & $\frac{1}{\sqrt{2}s_W}$ & 0
\\
\hline
$Z \bar{u} u$ & $\frac{-1+4c_W^2}{6c_Ws_W}$ & $\frac{-2s_W}{3c_W}$
\\
\hline
$Z \bar{d} d$ & $-\frac{1+2c_W^2}{6c_Ws_W}$ & $\frac{s_W}{3c_W}$
\\
\hline
$X^+ \bar{U} d$ & $-\frac{\ii}{\sqrt{2}s_W}$ & 0
\\
\hline
$X^+ \bar{u} d$ & $-\ii\delta_{u_i}^* \frac{1}{\sqrt{2}s_W}$ & 0
\\
\hline
$Z' \bar{u} u$ & $\frac{3+t_W^2}{6s_W\sqrt{3-t_W^2}}$ & $\frac{3+t_W^2}{6s_W\sqrt{3-t_W^2}}$
\\
\hline
$Z' \bar{d} d$ & $\frac{2t_W^2}{3s_W\sqrt{3-t_W^2}}$ & $-\frac{t_W^2}{3s_W\sqrt{3-t_W^2}}$
\\
\hline
\end{tabular}
\end{center}
\caption{Vertices ${\rm [VFF]}=\ii e (g_LP_L+g_RP_R)$ for the quark sector in the universal embedding 
entering in our calculation. \label{tab:VFF-u}}
\end{table}

\section{Lepton Flavor Violating Processes}
\label{LFVprocesses}
\subsection{General structure}

Here we summarize the expressions of the final branching ratios and conversion rates in terms of the different form factors we will obtain in the following sections.  In general we use the notation in \cite{delAguila:2008zu,delAguila:2010nv}.
  
\subsubsection{$\muegamma$ branching ratio}

The partial width for $\muegamma$ can be expressed in terms of the dipole form factors of the lepton flavor changing one-loop vertex as:
\begin{equation}
\Gamma(\muegamma) = \frac{\alpha}{2}m_{\mu}^3\big(|F_M^\gamma|^2+|F_E^\gamma|^2\big)\,.
\end{equation}
The branching ratio is then obtained by dividing by the total muon decay width which can be approximated by
\begin{equation}
  \Gamma(\mu\to\e\nu_\mu\bar{\nu}_e) = \frac{G_F^2m_{\mu}^5}{192\pi^3}\,, \quad G_F=\frac{\pi\alpha_W}{\sqrt{2}M_W^2}\,,\quad
\alpha_W=\frac{\alpha}{s_W^2}\,. \label{muegammabr}
\end{equation}

\subsubsection{$\mueee$ branching ratio}

The treatment of the decay width for the SLH is very similar to the one for LHT.  The only difference here is that we have additional penguin diagrams due to the exchange of $Z'$ gauge bosons. We define the amplitudes and form factors as follows:
\begin{eqnarray}
{\cal M}_{\gamma{\rm penguin}}&=&
\frac{e^2}{Q^2}
\bar u(p_1)\left[Q^2\gamma^\mu(A_1^L P_L+A_1^R P_R)+m_\mu\ii \sigma^{\mu\nu}Q_\nu(A_2^LP_L+A_2^RP_R)\right]u(p)\
 \nn\\ && \times\bar u(p_2)\gamma_\mu v(p_3)- (p_1\leftrightarrow p_2)\,,
\nn \\
{\cal M}_{Z{\rm penguin}}&=&
\frac{e^2}{M_Z^2}
\bar u(p_1)\left[\gamma^\mu(F_L P_L+F_R P_R)\right]u(p)\
\bar u(p_2)\left[\gamma_\mu(Z_L^e P_L+Z_R^e P_R)\right] v(p_3)
\nn\\ && - (p_1\leftrightarrow p_2)\,,
\nn \\
{\cal M}_{Z'{\rm penguin}}&=&
\frac{e^2}{M_Z'^2}
\bar u(p_1)\left[\gamma^\mu(F'_L P_L+F'_R P_R)\right]u(p)\
\bar u(p_2)\left[\gamma_\mu({Z'}_L^e P_L+{Z'}_R^e P_R)\right] v(p_3)
\nn\\ && - (p_1\leftrightarrow p_2)\,,
\nn \\
{\cal M}_{\rm boxes}&=&
\quad e^2 B_1^L\left[\bar u(p_1)\gamma^\mu P_L u(p)\right]
            \left[\bar u(p_2)\gamma_\mu P_L v(p_3)\right] \nn\\
&&+e^2 B_1^R\left[\bar u(p_1)\gamma^\mu P_R u(p)\right]
            \left[\bar u(p_2)\gamma_\mu P_R v(p_3)\right] \nn\\
&&+e^2 B_2^L\left\{\left[\bar u(p_1)\gamma^\mu P_L u(p)\right]
            \left[\bar u(p_2)\gamma_\mu P_R v(p_3)\right]
             - (p_1\leftrightarrow p_2) \right\} \nn\\
&&+e^2 B_2^R\left\{\left[\bar u(p_1)\gamma^\mu P_R u(p)\right]
            \left[\bar u(p_2)\gamma_\mu P_L v(p_3)\right]
             - (p_1\leftrightarrow p_2) \right\} \nn\\
&&+e^2 B_3^L\left\{\left[\bar u(p_1) P_L u(p)\right]
            \left[\bar u(p_2) P_L v(p_3)\right]
             - (p_1\leftrightarrow p_2) \right\} \nn\\
&&+e^2 B_3^R\left\{\left[\bar u(p_1) P_R u(p)\right]
            \left[\bar u(p_2) P_R v(p_3)\right]
             - (p_1\leftrightarrow p_2) \right\} \nn\\
&&+e^2 B_4^L\left\{\left[\bar u(p_1) \sigma^{\mu\nu}P_L u(p)\right]
            \left[\bar u(p_2)\sigma_{\mu\nu} P_L v(p_3)\right]
             - (p_1\leftrightarrow p_2) \right\} \nn\\
&&+e^2 B_4^R\left\{\left[\bar u(p_1) \sigma^{\mu\nu}P_R u(p)\right]
            \left[\bar u(p_2)\sigma_{\mu\nu} P_R v(p_3)\right]
             - (p_1\leftrightarrow p_2) \right\},\label{bfact}
\end{eqnarray}
\begin{eqnarray}
&&A_1^L=F_L^\gamma/Q^2\,, \
  A_1^R=F_R^\gamma/Q^2\, \
  A_2^L=-(F_M^\gamma+\ii F_E^\gamma)/m_\mu\,, \
  A_2^R=-(F_M^\gamma-\ii F_E^\gamma)/m_\mu\,, \nn\\
&&F_L=-F_L^Z\,, \quad
  F_R=-F_R^Z\,, \quad
  F'_L=-F_L^{Z'}\,, \quad
  F'_R=-F_R^{Z'}\,. \label{verff}
\end{eqnarray}
We can then use the following expression \cite{Hisano:1995cp,Arganda:2005ji} to obtain the partial width:
\begin{align}
\Gamma(&\mueee)=\frac{\alpha^2m_\mu^5}{32\pi}\bigg[
|A_1^L|^2+|A_1^R|^2-2(A_1^LA_2^{R*}+A_2^LA_1^{R*}+{\rm h.c.})
\nn\\ &
+(|A_2^L|^2+|A_2^R|^2)\left(\frac{16}{3}\ln\frac{m_\mu}{m_e}-\frac{22}{3}\right)
+\frac{1}{6}(|\hat B_1^L|^2+|\hat B_1^R|^2)
+\frac{1}{3}(|\hat B_2^L|^2+|\hat B_2^R|^2) \nn\\ &
+\frac{1}{24}(|B_3^L|^2+|B_3^R|^2)
+6(|B_4^L|^2+|B_4^R|^2)
-\frac{1}{2}(B_3^LB_4^{L*}+B_3^RB_4^{R*}+{\rm h.c.}) \nn\\ & +\frac{1}{3}(A_1^L\hat B_1^{L*}+A_1^R\hat B_1^{R*}+A_1^L\hat B_2^{L*}+A_1^R\hat B_2^{R*}+{\rm h.c.}) \nn\\ & -\frac{2}{3}(A_2^R\hat B_1^{L*}+A_2^L\hat B_1^{R*}+A_2^L\hat B_2^{R*}+A_2^R\hat B_2^{L*}+{\rm h.c.}) \nn\\ &
+\frac{1}{3}\big\{2(|F_{LL}|^2+|F_{RR}|^2)+|F_{LR}|^2+|F_{RL}|^2 \nn\\ &
+(\hat B_1^LF_{LL}^*+\hat B_1^RF_{RR}^*+\hat B_2^LF_{LR}^*+\hat B_2^RF_{RL}^*+{\rm h.c.})
+2(A_1^LF_{LL}^*+A_1^RF_{RR}^*+{\rm h.c.}) \nn\\ &
+(A_1^LF_{LR}^*+A_1^RF_{RL}^*+{\rm h.c.})
-4(A_2^RF_{LL}^*+A_2^LF_{RR}^*+{\rm h.c.}) \nn\\ &
-2(A_2^LF_{RL}^*+A_2^RF_{LR}^*+{\rm h.c.})\big\}\bigg]\,,
\label{widthmueee}
\end{align}
where
\bea
F_{LL}=\frac{F_LZ^{e}_L}{M_{Z}^2},\quad
F_{RR}=\frac{F_RZ^{e}_R}{M_{Z}^2}, \quad
F_{LR}=\frac{F_LZ^{e}_R}{M_{Z}^2}, \quad
F_{RL}=\frac{F_RZ^{e}_L}{M_{Z}^2}.
\eea
Some box form factors have been redefined to include the contributions from the ${\rm Z}'$ penguins:
\bea
B_1^L&\to& \hat B_1^L=B_1^L + 2F'_{LL}\, , \\
B_1^R&\to& \hat B_1^R=B_1^R + 2F'_{RR}\, ,\\
B_2^L&\to& \hat B_2^L=B_2^L + F'_{LR}\, ,\\
B_2^R&\to& \hat B_2^R=B_2^R + F'_{RL}\, ,
\eea
with
\bea
F'_{LL}=\frac{F'_LZ^{'e}_L}{M_{Z'}^2}\,,\quad
F'_{RR}=\frac{F'_RZ^{'e}_R}{M_{Z'}^2}\,, \quad
F'_{LR}=\frac{F'_LZ^{'e}_R}{M_{Z'}^2}\,, \quad
F'_{RL}=\frac{F'_RZ^{'e}_L}{M_{Z'}^2}\,.
\eea
In our case many of the form factors are zero so the expression above will be somewhat simplified.

\subsubsection{$\mueN$ conversion rate}

The $\mu\to\e$ conversion process \cite{Hisano:1995cp,Arganda:2005ji} is similar to $\mueee$ and differs only in that the lower part of the diagrams is coupled to quarks instead of leptons.  Also, we do not have identical particles in the final state.  We will sort the form factors as follows:
\begin{eqnarray}
\mathcal{M}_{\gamma \textrm{peng}}&=&-\frac{e^2}{Q^2}\bar{u}_{\rm e}(p_1)
\left[Q^2\gamma^\mu(A_1^LP_L+A_1^RP_R)+m_\mu i \sigma^{\mu\nu}Q_\nu(A_2^LP_L+A_2^RP_R)\right]u_{\mu}(p),\nonumber \\
&&\times \bar{u}_q(p_2)Q_q\gamma_\mu v_q(p_3)\,,\\
\mathcal{M}_{\rm Zpeng} &=& \frac{e^2}{M^2_Z}\bar{u}_{\rm e}(p_1)
\left[\gamma^\mu(F_LP_L+F_RP_R)\right]u_{\mu}(p)\times\bar{u}_q(p_2)\gamma_\mu \frac{Z_L^q+Z_R^q}{2}v_q(p_3)\,,\\
\mathcal{M}_{\rm {Z'}peng} &=& \frac{e^2}{M^2_{Z'}}\bar{u}_{\rm e}(p_1)
\left[\gamma^\mu(F'_LP_L+F'_RP_R)\right]u_{\mu}(p)\times\bar{u}_q(p_2)\gamma_\mu \frac{{Z'}_L^q+{Z'}_R^q}{2}v_q(p_3)\,,\\
\mathcal{M}_{\rm box}^q &=& e^2 \frac{B_{1q}^L}{2}\bar{u}_{\rm e} \gamma^\mu P_L u_\mu(p) \times
\bar{u}_q(p_2)\gamma_\mu v_q(p_3)\,.
\end{eqnarray}
We have already taken into accounto that, out of the original box form factors, only $B_1^L$ is non-zero.  The process width is \cite{Arganda:2005ji}:
\begin{equation}
\Gamma(\mu\to \textrm{e})=\alpha^5\frac{Z_{\rm eff}^4}{Z}|F(q)|^2m_{\mu}^5 
 \left| 2Z(A_1^L-A_2^R)-(2Z+N)\bar{B}_{1u}^L-(Z+2N)\bar{B}_{1d}^L\right|^2\,.
\end{equation}
The vertex form factors are as for $\mueee$ and were given in (\ref{verff}). We also have defined
\begin{equation}
\bar{B}_{1q}^L=B_{1q}^L+\frac{(Z_L^q+Z_R^q)F_L}{M_Z^2}+\frac{({Z'}_L^q+{Z'}_R^q)F'_L}{M_{Z'}^2}
\label{B1Lbar}
\end{equation}
to include the contributions from the $Z'$ penguins.

The conversion rate is obtained by dividing by the muon capture rate:
\begin{equation}
\mathcal{R}=\frac{\Gamma(\mu\to {\rm e})}{\Gamma_{\rm capt}}\, .
\end{equation}
The nuclei we will consider are ${}^{48}_{22}\textrm{Ti}$ and ${}^{197}_{\ 79}\textrm{Au}$, whose relevant parameters are listed in table~\ref{tab3}. 

\begin{table}
\begin{center}
\begin{tabular}{|c|rrccc|}
\hline
Nucleus & $Z$ & $N$ & $Z_{\rm eff}$ & $F(q)$ & $\Gamma_{\rm capt}$ [GeV] \\
\hline
${}^{48}_{22}\textrm{Ti}$ & 22 & 26 & 17.6 & 0.54 &  $1.7\times 10^{-18}$ \\
${}^{197}_{\ 79}\textrm{Au}$ & 79 & 118 & 33.5 & 0.16 & $8.6\times 10^{-18}$ \\
\hline
\end{tabular}
\end{center}
\caption{Relevant input parameters for the nuclei under study. From \cite{Kitano:2002mt}. \label{tab3}}
\end{table}

\subsection{$\muegamma$}

\begin{figure}
\begin{center}
\begin{tabular}{cccc}
\includegraphics[scale=0.6]{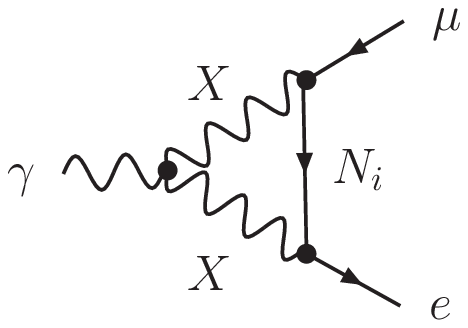} & \hspace{-12mm}
\includegraphics[scale=0.6]{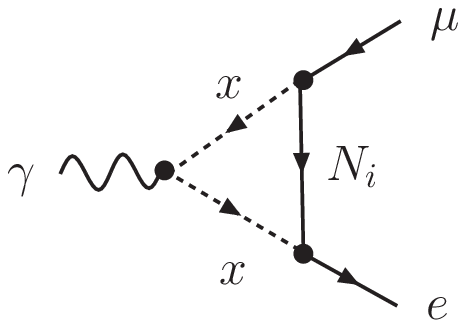} & \hspace{-12mm}
\includegraphics[scale=0.6]{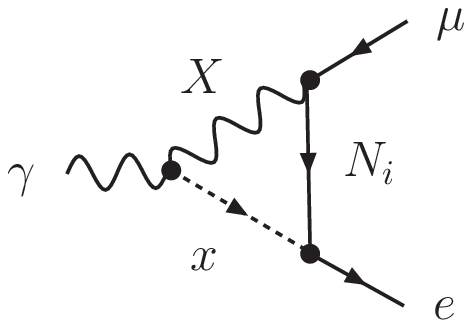} & \hspace{-12mm}
\includegraphics[scale=0.6]{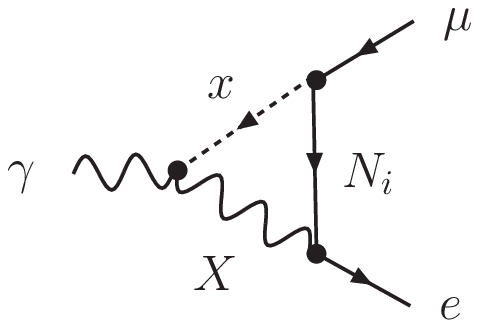} \\
\includegraphics[scale=0.6]{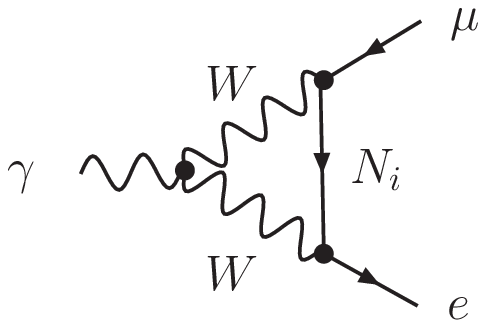} & \hspace{-12mm}
\includegraphics[scale=0.6]{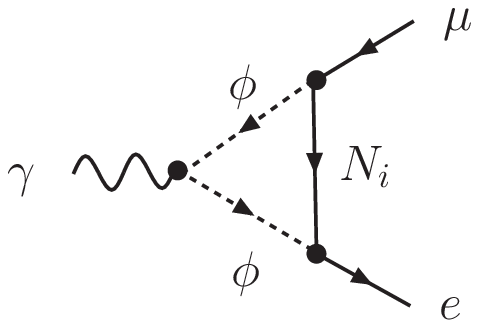} & \hspace{-12mm}
\includegraphics[scale=0.6]{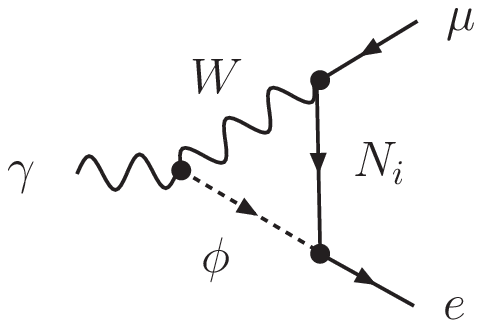} & \hspace{-12mm}
\includegraphics[scale=0.6]{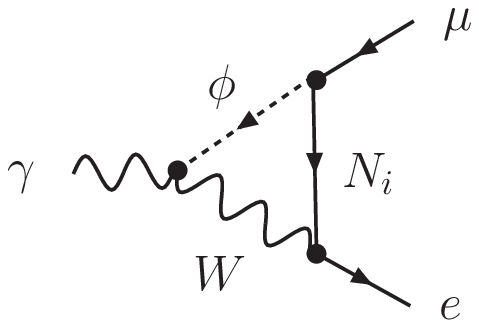} \\
II & IV & V & VI
\end{tabular}
\caption{Diagrams for $\muegamma$.\label{diagmuegamma}}
\end{center}
\end{figure}

In contrast to the LHT model, there is no symmetry preventing the coupling of two standard particles to a heavier one. Thus, we classify the contributions to $\muegamma$ into two types of diagrams (see figure~\ref{diagmuegamma}): those involving heavy $X$ gauge bosons and those with $W$ bosons in the loop. Since only dipole form factors contribute to this process, we have:
\bea
F_M^\gamma&=&F_M^\gamma|_X+F_M^\gamma|_W, \\
F_E^\gamma&=&F_E^\gamma|_X+F_E^\gamma|_W. \\
\eea

Defining the mass ratios:
\begin{equation}
x_i=\frac{m_{N_i}^2}{M_X^2}, \quad \omega = \frac{M_W^2}{M_X^2}
\end{equation}
we find the following contribution to the dipole form factors for the $X$-based diagrams:
\begin{eqnarray}
F_M^\gamma|_X=-\ii F_E^\gamma|_X = \frac{\alpha_W}{16\pi}\frac{m_\mu}{M_X^2}\sum_i
 V^{ie*} _\ell  V^{i\mu}_\ell    \ F_X(x_i)\,,
\end{eqnarray}
where we have introduced
\begin{eqnarray}
F_X(x) &=&M_1^2\left[
2\overline{C}_1-3\overline{C}_{11}-x\left(\overline{C}_0+3\overline{C}_1+\frac{3}{2}\overline{C}_{11}\right)\right]
\\
&=&
\frac{5}{6}-\frac{3x-15x^2-6x^3}{12(1-x)^3}+\frac{3x^3}{2(1-x)^4}\ln x.
\end{eqnarray}
The loop functions are summarized in Appendix~\ref{loopfunctions}.  In this case $Q^2=0$ for an on-shell photon.

For the $W$-based diagrams, we obtain
\begin{eqnarray}
F_M^\gamma|_W=-\ii F_E^\gamma|_W = \frac{\alpha_W}{16\pi} \delta_\nu^2 \frac{m_\mu}{M_W^2}\sum_i V^{ie*} _\ell  V^{i\mu}_\ell   \ F_W(x_i/\omega)\,,
\end{eqnarray}
where
\begin{eqnarray}
F_W(x) &=& M_1^2 \ x \left[ \overline{C}_0 + \overline{C}_1 - \frac{3}{2} \overline{C}_{11} \right] \\
& = & \frac{x(-7+5x+8x^2)}{12(1-x)^3}+\frac{x^2(-2+3x)}{2(1-x)^4}\ln{x}\,.
\end{eqnarray}

The total dipole form factors are therefore:
\begin{equation}
F_M^\gamma = -\ii F_E^\gamma = \frac{\alpha_W}{16\pi} \frac{m_\mu}{M_W^2} \sum_i V^{ie*} _\ell  V^{i\mu}_\ell  \left[\frac{v^2}{2f^2} \ F_X(x_i) + \delta_\nu^2 \ F_W(x_i/\omega) \right]\,.
\end{equation}

It should be noted, however, that in order to single out the relevant contributions in $W$ based diagrams, it is necessary to expand the loop functions with arguments $x_i/\omega$ as a series in $\omega\propto v^2/f^2$ to see whether the term is of leading order ($v^2/f^2$) or of higher order.  Nonetheless, once a term is accepted, we use the full expression for the numerical calculation. This is because we can not safely expand in orders of $\omega$ and then take a low $x_i$ limit since the limits $v^2/f^2 \rightarrow 0$ and $x_i \rightarrow 0$ do not commute.  Expanding in $\omega$ spoils the low $x_i$ behavior since $\omega = 0$ is a singular point. In this limit some terms in the series are divergent for low $x_i$ (for instance $\propto \ln{x_i}$) and give abnormally large contributions in this area not corresponding to the physical case, and it is therefore safer to keep the full expressions for the loop functions. We apply these criteria to all $W$ based contributions discussed in this paper.

\subsection{$\mueee$}

\begin{figure}[p]
\begin{center}
\includegraphics[scale=0.6]{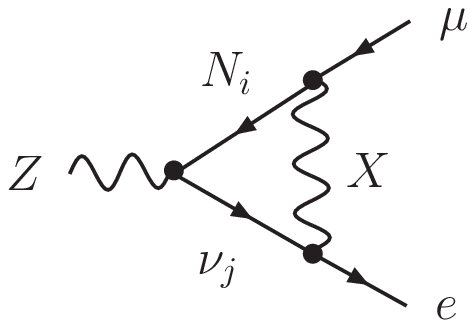}\hspace{-5mm} \includegraphics[scale=0.6]{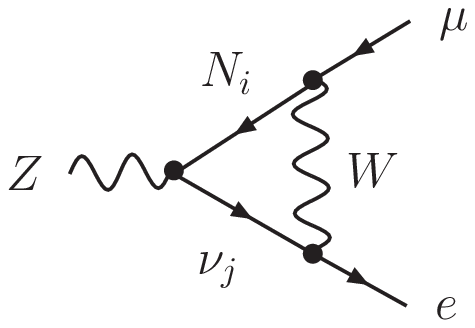} \hfill
\includegraphics[scale=0.6]{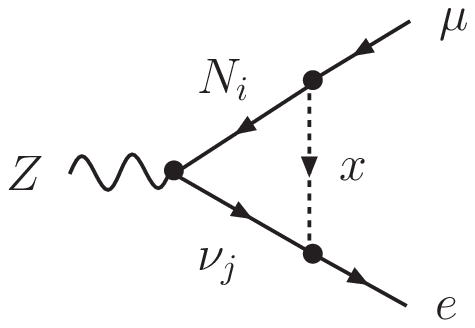}\hspace{-5mm} \includegraphics[scale=0.6]{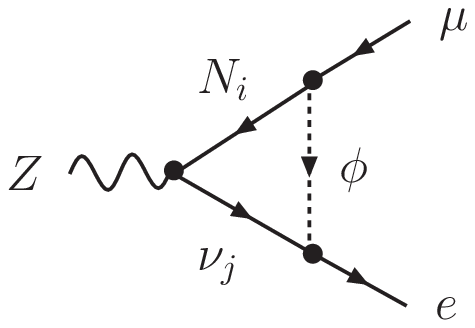}
\begin{tabular}{ccc}
\includegraphics[scale=0.6]{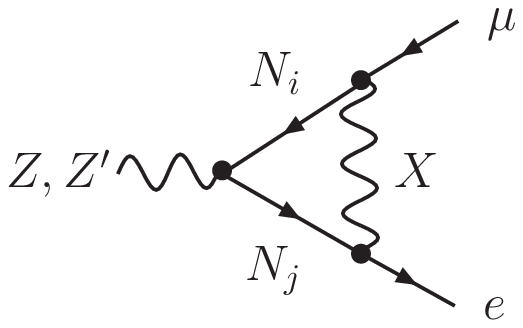} &
\includegraphics[scale=0.6]{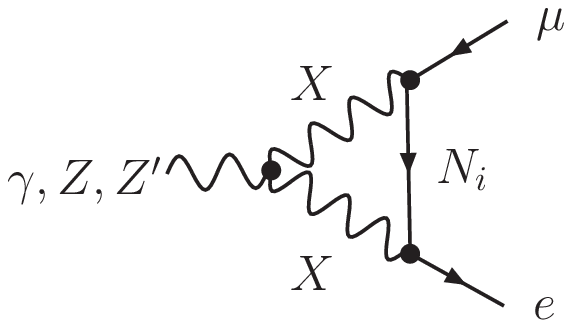} &
\includegraphics[scale=0.6]{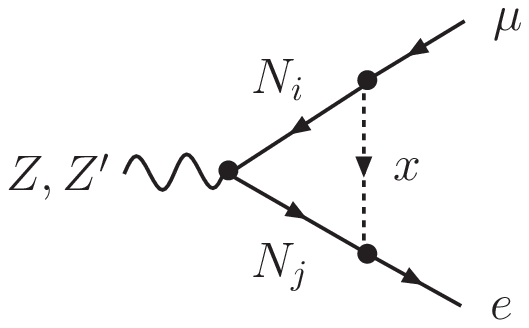} \\
\includegraphics[scale=0.6]{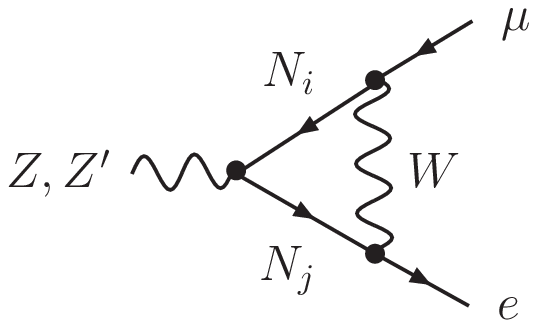} &
\includegraphics[scale=0.6]{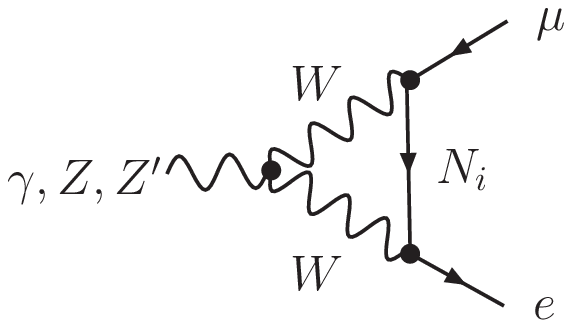} &
\includegraphics[scale=0.6]{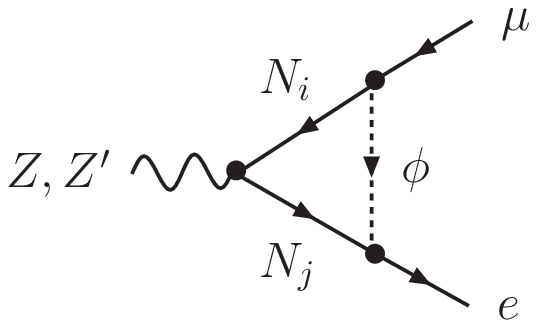} \\
I & II & III \\
\includegraphics[scale=0.6]{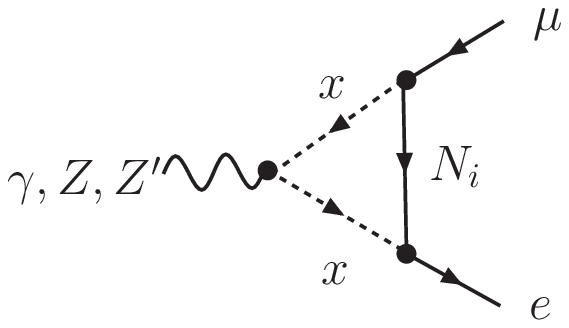} &
\includegraphics[scale=0.6]{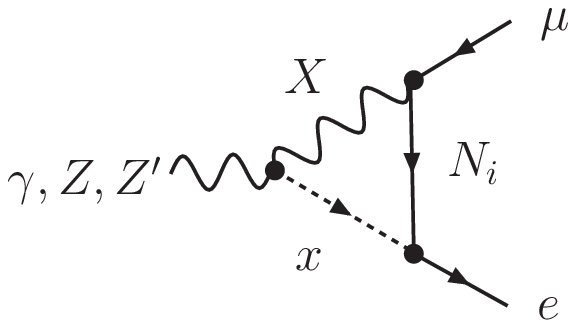} &
\includegraphics[scale=0.6]{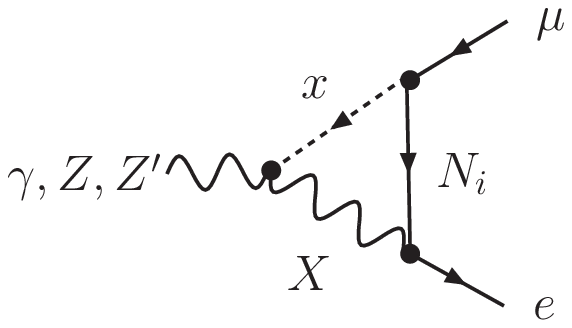} \\
\includegraphics[scale=0.6]{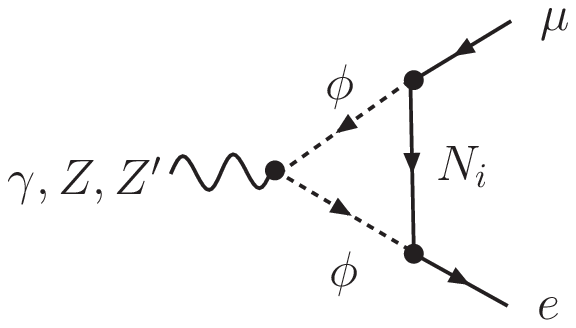} &
\includegraphics[scale=0.6]{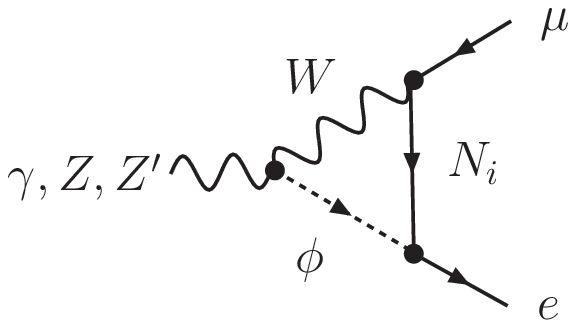} &
\includegraphics[scale=0.6]{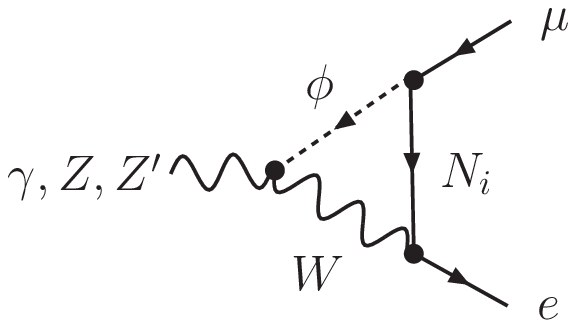} \\
IV & V & VI
\end{tabular}
\centerline{\includegraphics[scale=0.6]{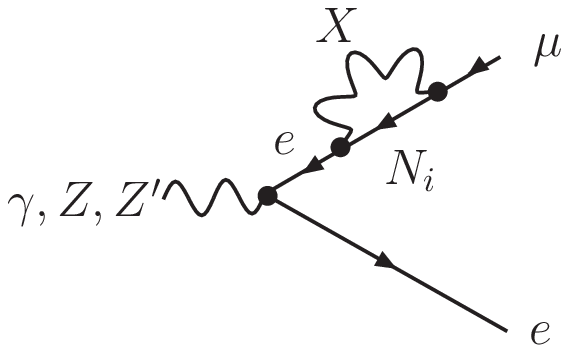} \hspace{-7mm}
\includegraphics[scale=0.6]{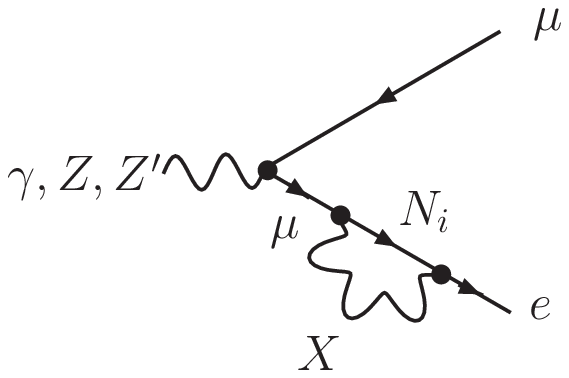} \hspace{-7mm}
\includegraphics[scale=0.6]{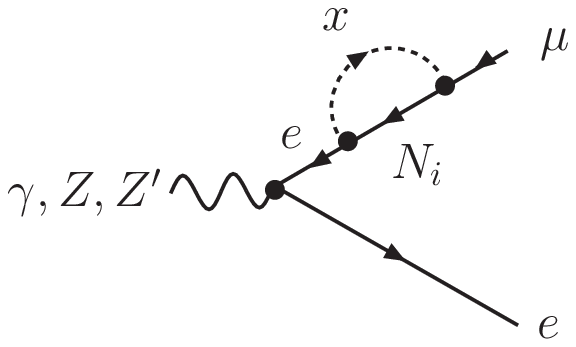} \hspace{-7mm}
\includegraphics[scale=0.6]{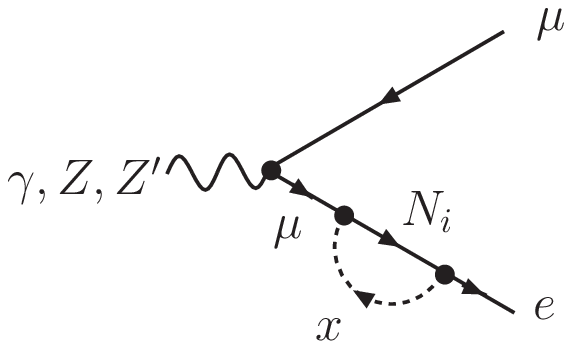}} 
\centerline{\includegraphics[scale=0.6]{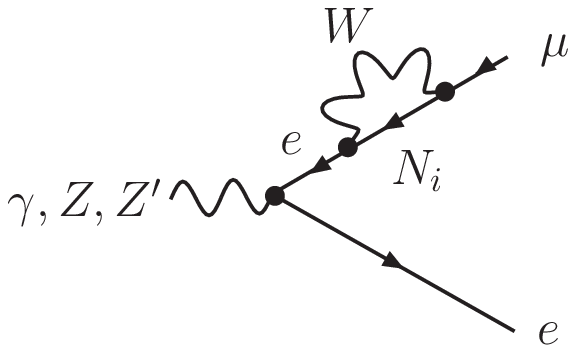} \hspace{-7mm}
\includegraphics[scale=0.6]{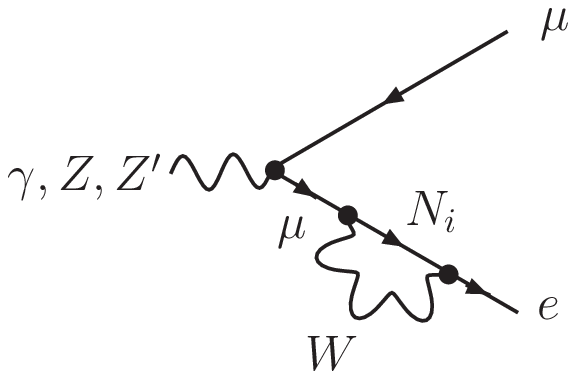} \hspace{-7mm}
\includegraphics[scale=0.6]{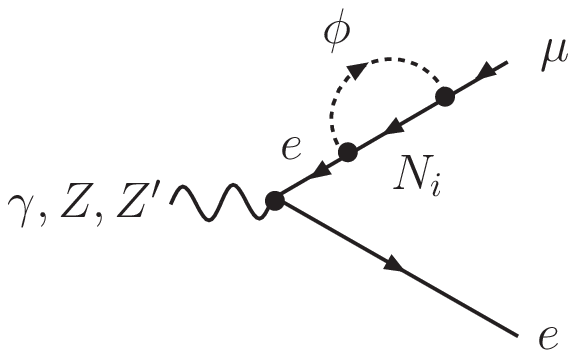} \hspace{-7mm}
\includegraphics[scale=0.6]{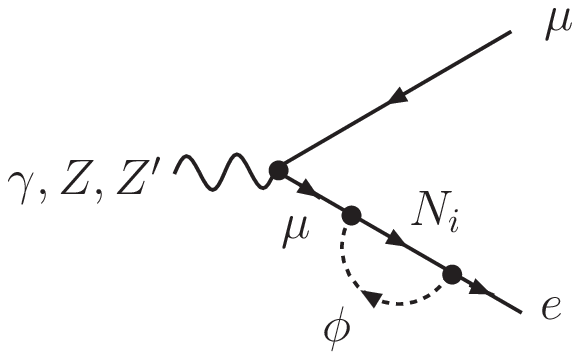}}
\caption{Relevant triangle and self-energy diagrams for $\mueee$.  Others give subleading contributions.}
\label{mueeediag}
\end{center}
\end{figure}

Here we have contributions from diagrams with $\gamma$, $Z$ and $Z'$ penguins as well as boxes.  The relevant diagrams are displayed in figure~\ref{mueeediag}.

\subsubsection{Photon penguins}

The dipole form factors are the same as in the $\muegamma$ case. For those terms we can set $Q^2=0$ as in $\muegamma$, since $Q^2$ is small in $\mueee$.  However, the $F_L$ and $F_R$ form factors require loop functions up to order $Q^2/M^2$ where $M$ is the mass of the gauge boson in the loop, because these terms need to cancel the $Q^{-2}$ factor coming from the photon propagator.

In $F_L$ and $F_R$ we neglect terms of order $m_\ell^2/M^2$.  This means that, $F_R \simeq 0$.  We again divide the contributions into two groups.
Firstly, the $X$ related diagramas work out as
\begin{align}
F_L^\gamma|_X= \frac{\alpha_W}{4\pi}\sum_i  V^{ie*}_\ell V^{i\mu}_\ell   \ G_X(x_i),
\end{align}
where
\bea
G_X(x)&=&\!\!(1-\delta_\nu^2)\left[-\frac{1}{2}+\overline B_1+6\overline C_{00} + x\left(\frac{1}{2}\overline B_1+\overline C_{00}-M_{X}^2\overline C_0\right)\right]
-\left(2\overline C_1+\frac{1}{2}\overline C_{11}\right)Q^2
\nn\\
&=&(1-\delta_\nu^2)\left(\Delta_\epsilon-\ln\frac{M_{X}^2}{\mu^2}\right)+\frac{Q^2}{M_{X}^2}G^{(1)}_X(x)
+{\cal O}\left(\frac{Q^4}{M_X^4}\right)\,,
\\
G^{(1)}_X(x)&=&
-\frac{5}{18}
+\frac{x(12+x-7x^2)}{24(1-x)^3}+\frac{x^2(12-10x+x^2)}{12(1-x)^4}\ln x\,,
\eea
and $\Delta_\epsilon=\frac{2}{\epsilon}-\gamma+\ln 4\pi$, divergent in four dimensions.
Therefore:
\begin{equation}
F_L^\gamma|_X= \frac{\alpha_W}{4\pi}\frac{Q^2}{M_{X}^2}\sum_i  V^{ie*} _\ell V^{i\mu}_\ell   \ G_X^{(1)}(x_i)\,.
\end{equation}

The $W$-based diagrams contribute with:
\begin{eqnarray}
F_L^\gamma|_W= \frac{\alpha_W}{4\pi}\delta_\nu^2\sum_i  V^{ie*} _\ell V^{i\mu}_\ell   \ G_W(x_i/\omega)\,,
\end{eqnarray}
where
\bea
G_W(x)&=&\!\!-\frac{1}{2}+\overline B_1+6\overline C_{00} + x\left(\frac{1}{2}\overline B_1+\overline C_{00}-M_1^2\overline C_0\right)
\nn\\
&=&\Delta_\epsilon-\ln\frac{M_{W}^2}{\mu^2}+\frac{Q^2}{M_{W}^2}G^{(1)}_W(x)
+{\cal O}\left(\frac{Q^4}{M_W^4}\right)\,,
\\
G^{(1)}_W(x)&=& \frac{1}{6} - \frac{x(-2+7x-11x^2)}{72(1-x)^3} + \frac{x^4}{12(1-x)^4}\ln{x}\, .
\eea
That is:
\begin{equation}
F_L^\gamma|_W= \frac{\alpha_W}{4\pi}\frac{Q^2}{M_{W}^2}\delta_\nu^2\sum_i  V^{ie*} _\ell V^{i\mu}_\ell   \ G_W^{(1)}(x_i/\omega)\,.
\end{equation}

\subsubsection{$Z$ penguins}

There are three pieces in this section: two of them ($F_L^Z|_X$ and $F_L^Z|_W$) involve only heavy neutrinos in the loop, and the third ($F_L^Z|_{\rm hl}$) contains one heavy and one light neutrino exchanging either a $W$ or an $X$ boson,
\begin{equation}
F_L^Z = F_L^Z|_X + F_L^Z|_W + F_L^Z|_{\rm hl}\, .
\end{equation}
Again $F_R \simeq 0$ if we neglect $m_\ell^2/M^2$.
The $X$-based diagrams result in:
\begin{eqnarray*}
F_L^Z|_{X}&=& \frac{\alpha_W}{4\pi}\frac{1}{c_Ws_W}\sum_i V^{ie*}_\ell V^{i\mu}_\ell\ \left(\frac{c_W\delta_Z}{\sqrt{3-t_W^2}}\ I_X(x_i) + \delta_\nu^2 \ H_X(x_i) \right)\, ,
\end{eqnarray*}
where
\begin{eqnarray}
I_X(x) & = & -\frac{6x-x^2}{2(1-x)}-\frac{2x+3x^2}{2(1-x)^2}\ln{x} , \label{I_X}\\
H_X(x) & = & \frac{x}{4}+\frac{x}{2(1-x)}\ln{x}\,.
\end{eqnarray}

The $W$ boson diagrams give the following contribution:
\begin{eqnarray*}
F_L^Z|_W &=& \frac{\alpha_W}{4\pi}\frac{1}{s_Wc_W}\ \delta_\nu^2 \ \sum_i  V^{ie*} _\ell V^{i\mu}_\ell   \left[ H_W(x_i/\omega) - \frac{2 +(1-t_W^2)t_\beta}{8} \ \delta_\nu^2 \ I_W(x_i/\omega)\right]\,,
\end{eqnarray*}
where
\begin{eqnarray}
 H_W(x) & = & \frac{1}{8} + \frac{5x}{4(1-x)}+\frac{5x^2}{4(1-x)^2}\ln{x} ,\\
 I_W(x) & = & \frac{x^2}{1-x} + \frac{x^2}{(1-x)^2}\ln{x}\, . \label{I_W}
\end{eqnarray}

Finally, diagrams where the $Z$ couples one heavy to one light neutrino contribute with:
\begin{eqnarray}
F_L^Z|_{\rm hl} & = & \frac{\alpha_W}{2\pi}\frac{1}{s_Wc_W} \ \delta_\nu^2 \ \sum_i  V^{ie*} _\ell V^{i\mu}_\ell   \left\{ \hat{C}_{00}(M_W^2,0;x_i/\omega) - \hat{C}_{00}(M_X^2,0;x_i) \right\} \\
& = & \frac{\alpha_W}{4\pi}\frac{1}{s_Wc_W} \ \delta_\nu^2 \ \sum_i  V^{ie*} _\ell V^{i\mu}_\ell  
\left[ H_Z(x_i/\omega) - H_Z(x_i) \right]\,,
\end{eqnarray}
where
\begin{equation}
H_Z(x) = \frac{x\ln{x}}{2(1-x)}\, .
\end{equation}

\subsubsection{$Z'$ penguins}

Here we have two contributions:
\begin{equation}
F_L^{Z'} = F_L^{Z'}|_X + F_L^{Z'}|_W\,.
\end{equation}
There is no piece analogous to $F_L^Z|_{\rm hl}$ since the $Z'$ has an additional $v^2/f^2$ from its propagator that makes those terms subleading. The form factors read:
\begin{eqnarray}
F_L^{Z'}|_X & = & \frac{\alpha_W}{4\pi}\frac{1}{s_W\sqrt{3-t_W^2}}
\sum_i  V^{ie*} _\ell V^{i\mu}_\ell I_X(x_i)\,, \\
F_L^{Z'}|_W & = & \frac{\alpha_W}{8\pi} \frac{\delta_\nu^2 }{s_W\sqrt{3-t_W^2}}
\sum_i  V^{ie*} _\ell V^{i\mu}_\ell I_W(x_i/\omega)\,.
\end{eqnarray}
where $I_X$ and $I_W$ are defined in (\ref{I_X}) and (\ref{I_W}).

\subsubsection{Box contributions}

\begin{figure}
\begin{center}
\begin{tabular}{cccc}
\includegraphics[scale=0.6]{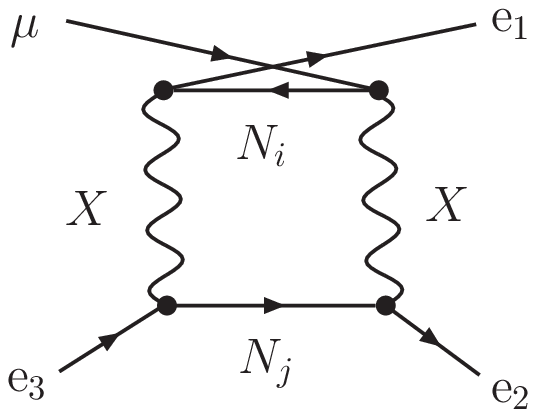} & \hspace{-8mm}
\includegraphics[scale=0.6]{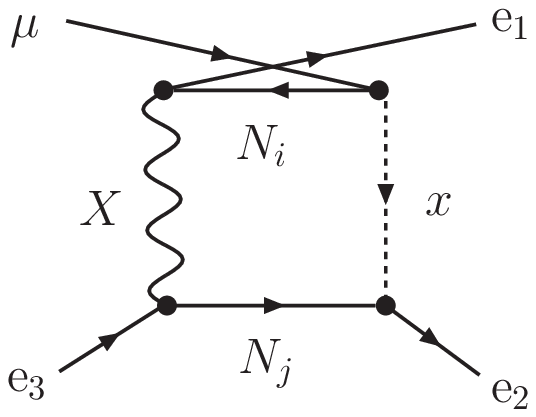} & \hspace{-8mm}
\includegraphics[scale=0.6]{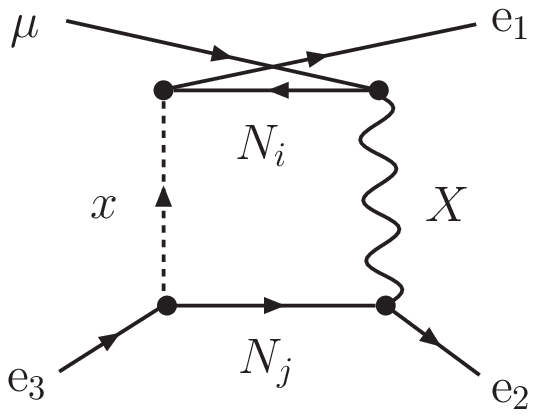} & \hspace{-8mm}
\includegraphics[scale=0.6]{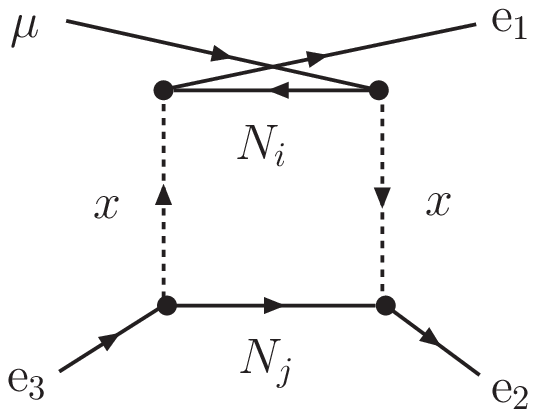}\\
\includegraphics[scale=0.6]{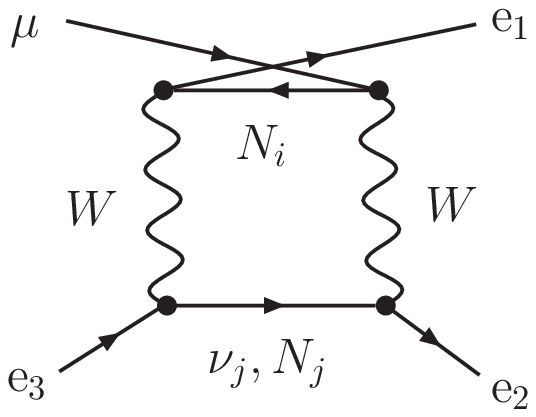} & \hspace{-8mm}
\includegraphics[scale=0.6]{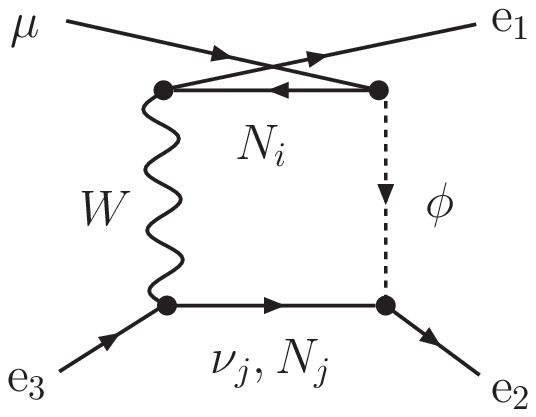} & \hspace{-8mm}
\includegraphics[scale=0.6]{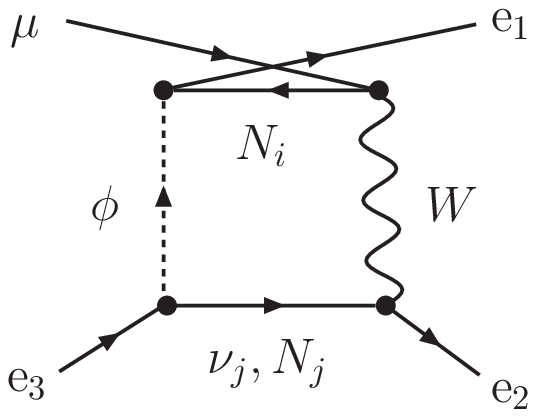} & \hspace{-8mm}
\includegraphics[scale=0.6]{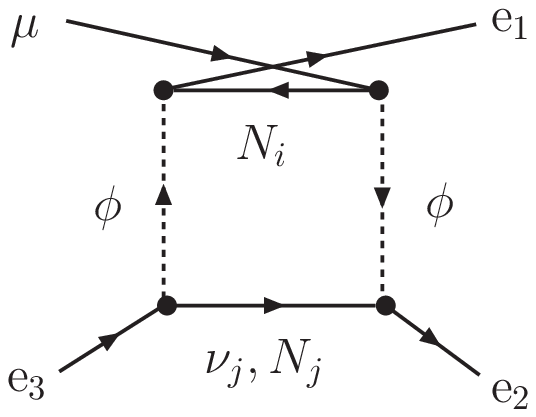}\\
\includegraphics[scale=0.6]{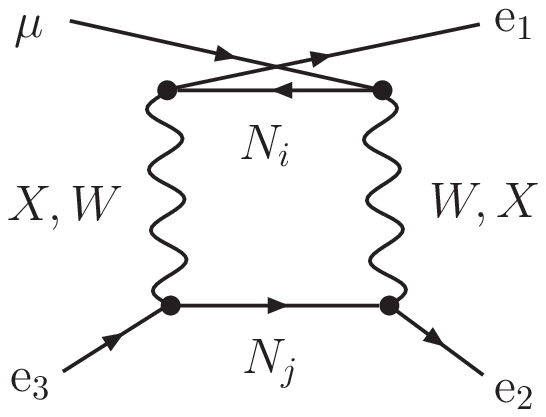} & \hspace{-8mm}
\includegraphics[scale=0.6]{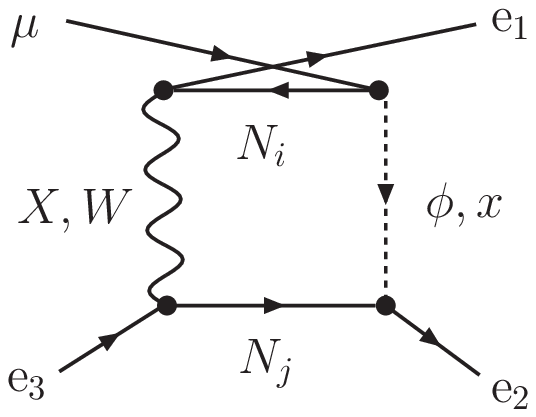} & \hspace{-8mm}
\includegraphics[scale=0.6]{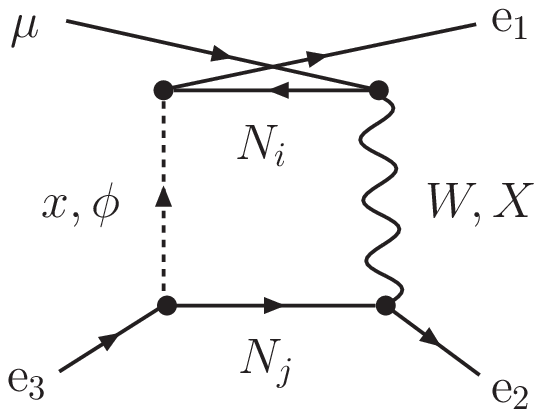} & \hspace{-8mm}
\includegraphics[scale=0.6]{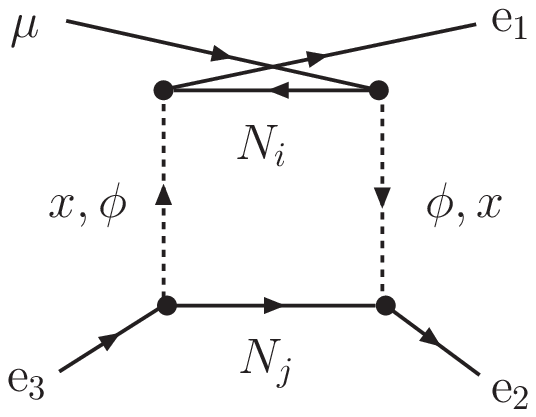}
\end{tabular}
\caption{Box diagrams for $\mueee$. \label{fig:box}}
\end{center}
\end{figure}

Only $W$ and $X$ particles can be involved in the loop (see figure~\ref{fig:box}).  Crossed diagrams, not shown in the figure, contribute a factor 2 due to Fierz identities \cite{Nishi:2004st}.  Neglecting ${m_\ell}/{M}$ we have contributions only to the $B^L_1$ form factor, divided in three terms:
\begin{equation}
B_1^L = B_1^L|_X + B_1^L|_W + B_1^L|_{WX}\,,
\end{equation}
where
\begin{eqnarray}
B_1^L|_X & = & \frac{\alpha_W}{8\pi}\frac{1}{s_W^2}\frac{1}{M_X^2}\sum_{ij} \chi_{ij} \Bigg[ \left(1+\frac{1}{4}x_ix_j\right)\widetilde{d}_0(x_i,x_j) - 2x_ix_jd_0(x_i,x_j)\Bigg], \nonumber \\
& & \\
B_1^L|_W & = & \frac{\alpha_W}{8\pi}\frac{1}{s_W^2}\frac{\delta_\nu^4}{M_W^2}
\sum_{ij} \left[\chi_{ij} \frac{x_ix_j}{4\omega^2}\tilde{d}_0(x_i/\omega,x_j/\omega)
\right]\,,\\
B_1^L|_{WX} & = & \frac{\alpha_W}{8\pi}\frac{1}{s_W^2}\frac{\delta_\nu^2}{M_W^2}
\sum_{ij} \chi_{ij} x_ix_j\left[\frac{1}{2}\widetilde{d}_0(\omega,x_i,x_j) - 2d_0(\omega,x_i,x_j)\right]\,,
\end{eqnarray}
and 
\bea
\chi_{ij} =  V^{ie*}_\ell V^{i\mu}_\ell  |V^{je}_\ell|^2\,.
\eea

\subsection{$\mue$ conversion in nuclei}

\begin{figure}
\begin{center}
\begin{tabular}{cccc}
\includegraphics[scale=0.6]{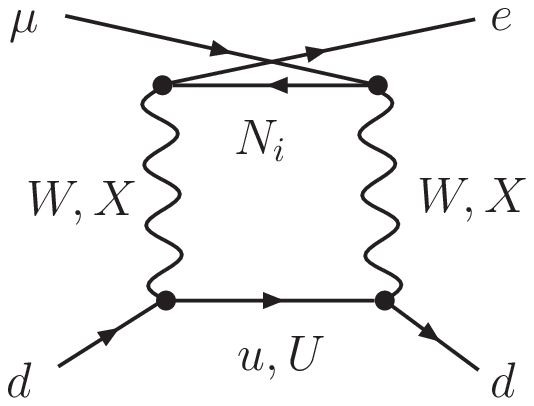} & \hspace{-8mm}
\includegraphics[scale=0.6]{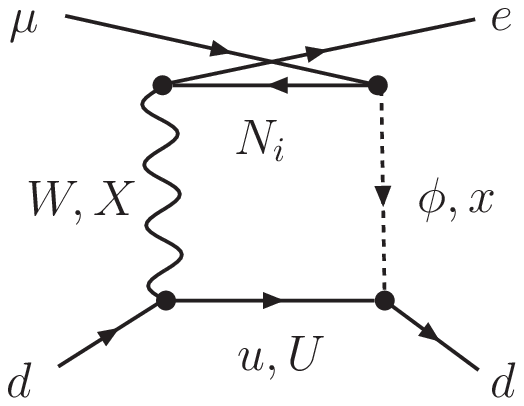} & \hspace{-8mm}
\includegraphics[scale=0.6]{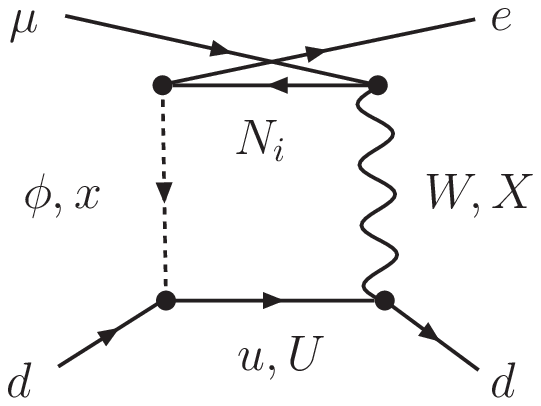} & \hspace{-8mm}
\includegraphics[scale=0.6]{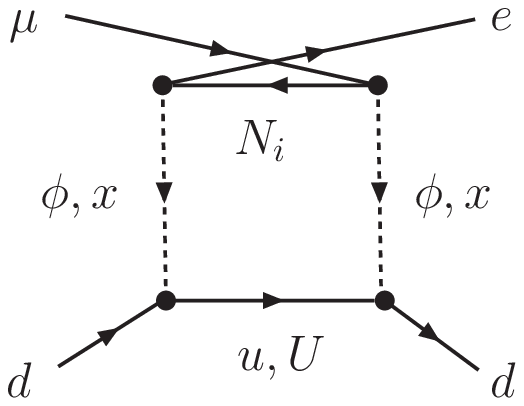} \\
\includegraphics[scale=0.6]{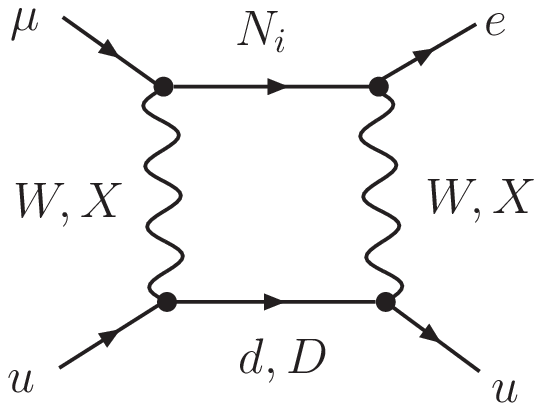} & \hspace{-8mm}
\includegraphics[scale=0.6]{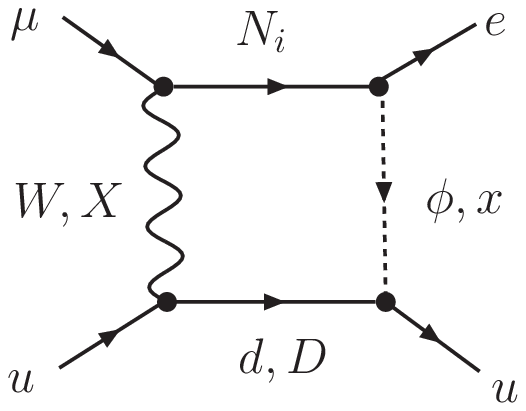} & \hspace{-8mm}
\includegraphics[scale=0.6]{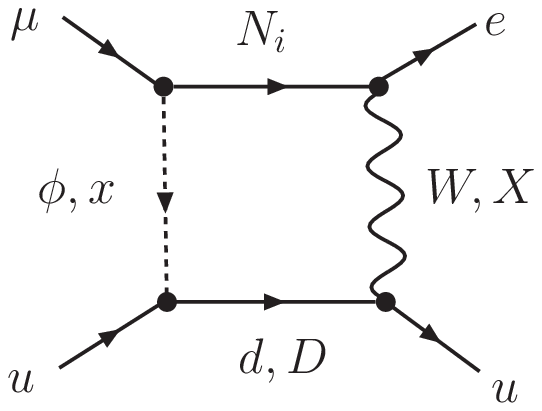} & \hspace{-8mm}
\includegraphics[scale=0.6]{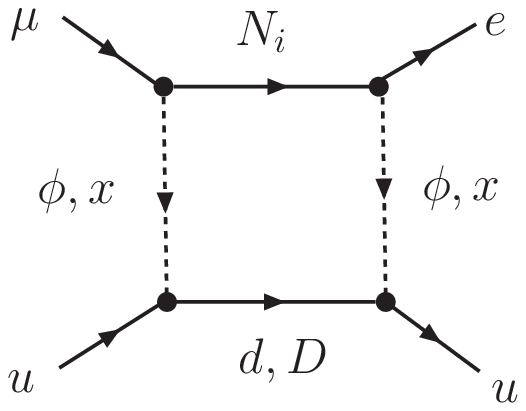}
\end{tabular}
\end{center}
\caption{Box diagrams for $\mueN$.\label{fig:boxconv}}
\end{figure}

The triangle form factors are the same as in the $\mueee$ process.
Only the box form factors need to be recalculated.  These are of course embedding dependent.  The diagrams include all combinations of quarks and gauge bosons (figure~\ref{fig:boxconv}).  We have assumed heavy quark degeneracy to suppress any mixing effects from this sector, which were not included in the Feynman rules anyway.  In this approximation, only diagrams with a $D$ quark appear in the anomaly-free embedding while only diagrams with a $U$ quark are included for the universal embedding.  Diagrams with light quarks appear in both embeddings but will be found to be a subleading contribution.  The form factors $B_{1q}^L$ refer to diagramas where the $q$ quark enters on the lower line.

In the anomaly-free embedding we obtain:
\begin{eqnarray}
B_{1u}^L & = &  -\frac{\alpha_W}{16\pi} \frac{1}{s_W^2}\frac{1}{M_W^2} \sum_i V^{ie*}_\ell V^{i\mu}_\ell \nonumber \Bigg\{ \frac{v^2}{2f^2} \bigg[ \left(4+\frac{1}{4}x_ix_D\right) \ \widetilde{d}_0(x_i,x_D) \\ 
&&- 2 x_ix_D \ d_0(x_i,x_D) \bigg] + \frac{|\delta_d|^2\delta_\nu^2}{4\omega^2} \ x_i x_D \ \widetilde{d}_0(x_i/\omega,x_D/\omega) \nonumber \\
&& + \delta_\nu (\delta_d +\delta_d^*) x_i x_D \bigg[ d_0(\omega,x_i,x_D) - \frac{1}{4} \ \widetilde{d}_0(\omega,x_i,x_D) \bigg] \Bigg\}\,, \\
B_{1d}^L & = & 0\,,
\end{eqnarray}
and in the universal embedding:
\begin{eqnarray}
B_{1u}^L & = & 0\,, \\
B_{1d}^L & = &  \frac{\alpha_W}{16\pi} \frac{1}{s_W^2}\frac{1}{M_W^2} \sum_i V^{ie*}_\ell V^{i\mu}_\ell
\Bigg\{ \frac{v^2}{2f^2} \bigg[ \left(1+\frac{1}{4}x_ix_U\right) \ \widetilde{d}_0(x_i,x_U) \\
&& - 2 x_ix_U \ d_0(x_i,x_U) \bigg] + \frac{|\delta_u|^2\delta_\nu^2}{4\omega^2} x_i x_U \ \widetilde{d}_0(x_i/\omega,x_U/\omega) \\
&& - \delta_\nu (\delta_u +\delta_u^*) x_i x_U \bigg[ d_0(\omega,x_i,x_U) - \frac{1}{4} \ \widetilde{d}_0(\omega,x_i,x_U) \bigg] \Bigg\}\,.
\end{eqnarray}
with 
\bea
x_D = m_D^2/M_X^2,\quad 
x_U = m_U^2/M_X^2.
\eea  
Here, the $Z'$ couplings ${Z'}_{L,R}^q$ appearing in (\ref{B1Lbar}) are also embedding dependent.

\section{Numerical results}
\label{num}
We now give the results for the different processes.  We follow closely the analysis in \cite{delAguila:2008zu,delAguila:2010nv} and restrict ourselves to the case of two lepton families to illustrate the physical behavior of the amplitudes.  This leaves us with four basic parameters:
\begin{itemize}

\item
The masses of the two heavy leptons, $m_{N_{1,2}}$, parameterized in terms of an average mass and a relative mass splitting:
\bea
\tilde{x} = \sqrt{x_1x_2}, \quad x_i=\frac{m_{N_i}^2}{M_X^2},\quad  \delta = \frac{m_{N_2}^2-m_{N_1}^2}{m_{N_1}m_{N_2}}.
\eea

\item
The only remaining mixing angle $\theta$ in $V_\ell$:
\begin{equation}
 V_\ell = \left( \begin{array}{cc} \cos \theta & \sin \theta \\ -\sin \theta & \cos \theta \end{array} \right).
\end{equation}

\item 
The Little Higgs breaking scale $f$.
\end{itemize}

Values in the theory for these parameters which may be considered \emph{natural} 
are $f\sim 1$ TeV, $\tilde{x} = \delta = 1$ and $\sin 2\theta = 1$.  We take those as reference values for our analysis as these are the typical values one could expect for this model.  There are also two additional parameters involved:  $t_\beta$ and the mass of the heavy quarks that appear in the box diagrams of the $\mueN$ conversion process.  We will parameterize the heavy quark masses
as $x_U = m_U^2/M_X^2$ and $x_D = m_D^2/M_X^2$ and will take reference values $x_U = x_D = 1$.  In order to comply with the limit $\delta_\nu<0.05$ we need to impose $ft_\beta \gtrsim 3.5$ TeV.  Therefore, to allow for values of $f$ below 1 TeV, we choose a reference value $t_\beta = 5$.

For the processes $\muegamma$ and $\mue$ conversion, all the form factors have the following general form:
\begin{equation}
 \mathcal{A} = \sum_{i=1,2} V^{ie*}_\ell V^{i\mu}_\ell F(x_i) = \frac{\sin 2\theta}{2} [F(x_1) - F(x_2)]\,.
\label{vertexdep}
\end{equation}
The branching ratio or conversion rate can then be approximated by:
\begin{equation}
 \mathcal{B} \sim \left|\frac{v^2}{f^2}\delta \sin{2\theta} \right|^2\, . \label{dep}
\end{equation}
The dependence on $\sin{2\theta}$ is exact while the dependence on $\delta$ is valid for small values of this parameter.  The dependence on $f$ is also approximate due to the fact that some loop functions depend on $x_i/\omega$.  However, the leading order in every form factor is $v^2/f^2$ and the $1/f^4$ behavior is a good approximation.  The dependence on $\tilde{x}$ cannot be expressed as simply.

For the $\mueee$ process, the vertex form factors are as in (\ref{vertexdep}) while the box form factors take the form:
\begin{equation}
\begin{split}
 \mathcal{A}= &\sum_{ij=1,2} V^{ie*}_\ell V^{i\mu}_{\ell} |V_{je}^\ell|^2 F(m_{N_i}^2,m_{N_j}^2) \\
=& -\frac{\sin 2\theta}{2} \Big(\cos^2 \theta \left[F(m_{N_1}^2,m_{N_1}^2) - F(m_{N_2}^2,m_{N_1}^2) \right] +\sin^2 \theta \left[F(m_{N_1}^2,m_{N_2}^2) - F(m_{N_2}^2,m_{N_2}^2) \right]\,.
\end{split}
\end{equation}
This makes the angle dependence in this process slightly different from the other cases.  However, the effect is small and the general behavior remains similar to (\ref{dep}).

\begin{table}[t]
\begin{center}
\begin{tabular}{|rcccccc|}
\hline
& $\muegamma$ & 
  $\mueee$  &   \multicolumn{2}{c}{$\mu\, {\rm Au}\to\e\, {\rm Au}$} &
  \multicolumn{2}{c|}{$\mu\, {\rm Ti}\to\e\, {\rm Ti}$} \\
\cline{2-7}
\multicolumn{1}{|c}{Limit}
& $1.2\times10^{-11}$  & $10^{-12}$ &    \multicolumn{2}{c}{$7\times10^{-13}$} & \multicolumn{2}{c|}{$4.3\times10^{-12}$} 
\\
\hline
$f/\mbox{TeV}>$ & 4.3 & 4.1 & 13.9 & (16.5) & 8.8 & (10.3) \\
$\sin2\theta<$  & 0.052 & 0.055 & 0.005 & (0.004) & 0.013 & (0.009) \\
$|\delta|<$     & 0.050 & 0.059 & 0.005 & (0.003) & 0.013 & (0.009) \\
\hline
\end{tabular}
\end{center}
\caption{Bounds on SLH parameters from present \cite{Brooks:1999pu,Bellgardt:1987du,Bertl:2006up,Dohmen:1993mp} limits on LFV processes. For $\mueN$ the numbers correspond to the anomaly-free (universal) embedding.}
\label{limits1}
\begin{center}
\begin{tabular}{|rcccc|}
\hline
& $\muegamma$ & 
  $\mueee$  &  \multicolumn{2}{c|}{$\mu\, {\rm Ti}\to\e\, {\rm Ti}$} \\
\cline{2-5}
\multicolumn{1}{|c}{Limit}
& $10^{-13}$ & $10^{-14}$ & \multicolumn{2}{c|}{$10^{-18}$}
\\
\hline
$f/\mbox{TeV}>$ & 14.2 & 12.9 & 397 & (468) \\
$\sin2\theta<$  & 0.004 & 0.005 & $<10^{-4}$ & ($<10^{-4}$) \\
$|\delta|<$     & 0.005 & 0.006 & $<10^{-4}$ & ($<10^{-4}$) \\
\hline
\end{tabular}
\end{center}
\caption{Bounds on SLH parameters from future \cite{Mori:2007zza,Adam:2009ci,Kuno:2005mm,Pasternak:2010zz} limits on LFV processes.}
\label{limits2}
\end{table}

\begin{figure}
\centering
  \begin{tabular}{cl}
\includegraphics[height=6cm]{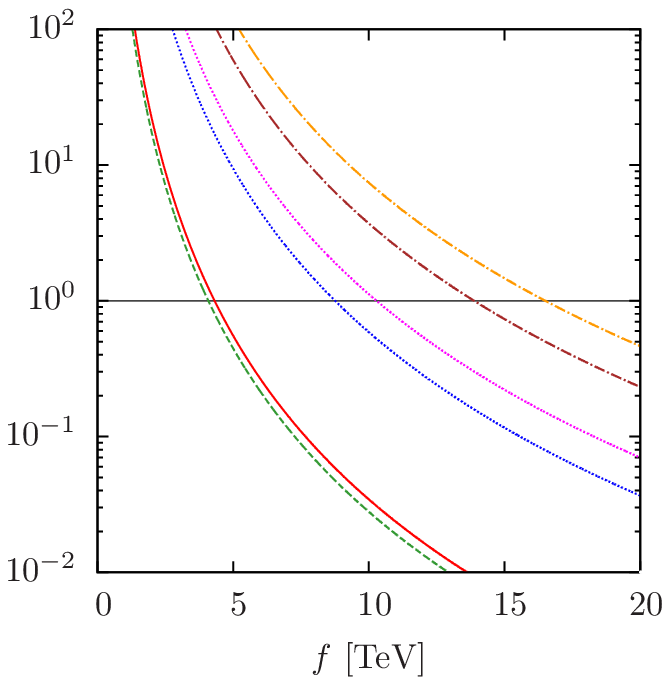} & \!\!\!\!\includegraphics[height=6cm]{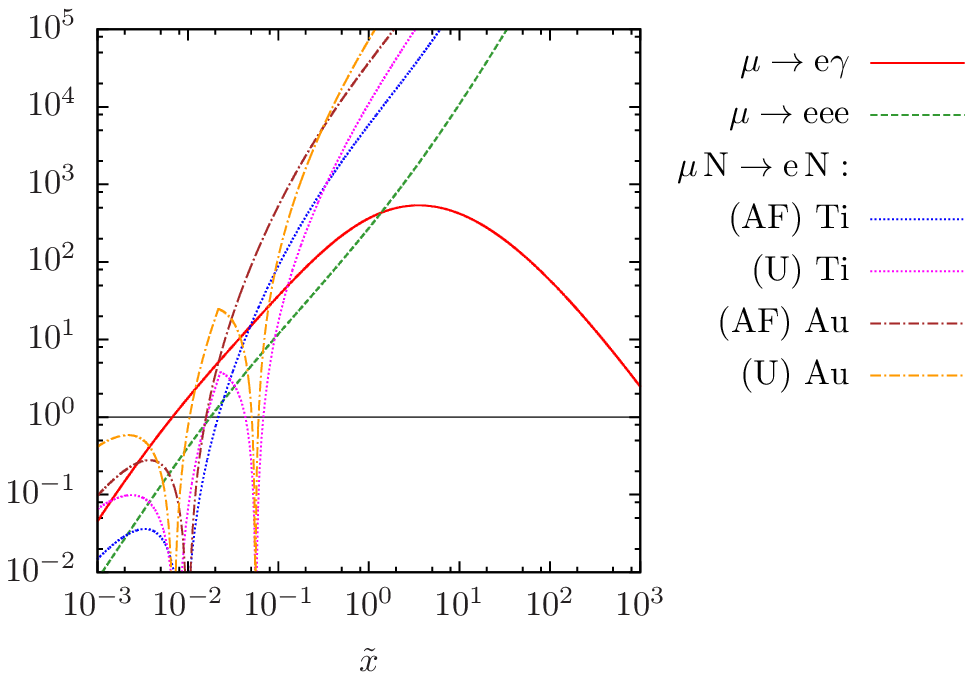} \\
\includegraphics[height=6cm]{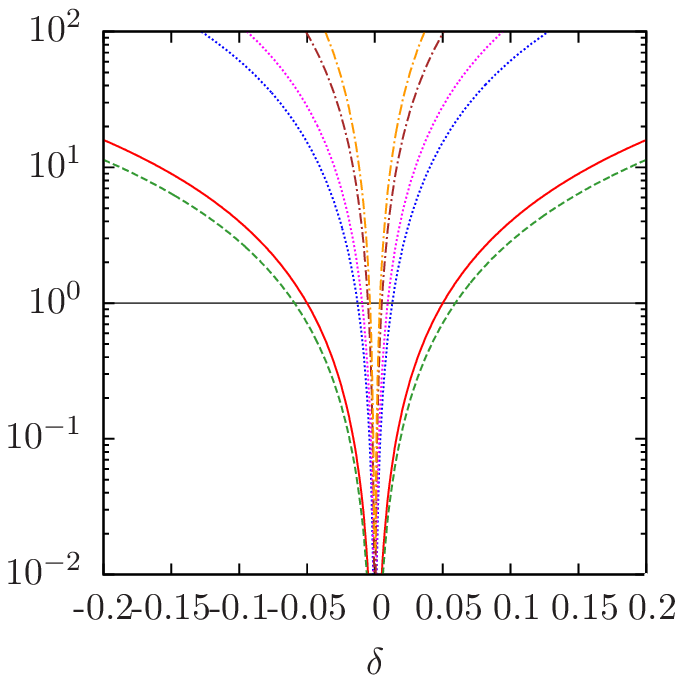} & \!\!\!\!\includegraphics[height=6cm]{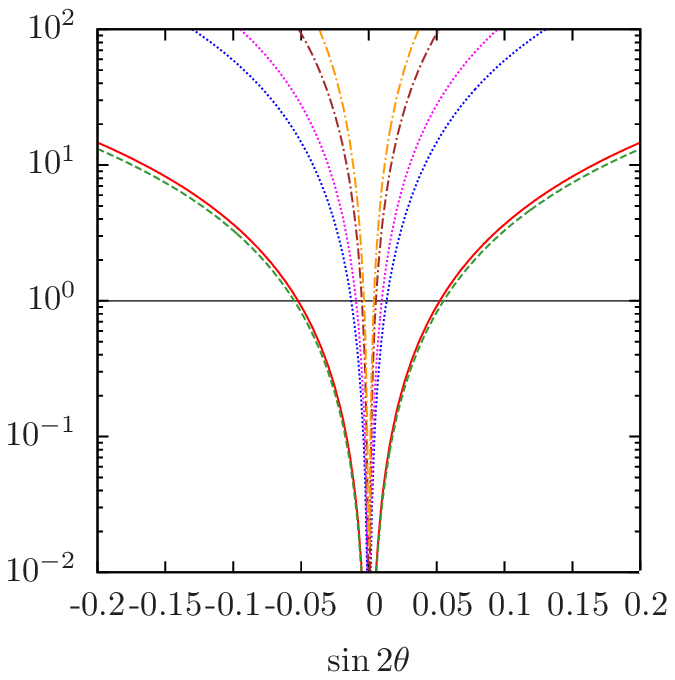}
 \end{tabular}
\caption{Ratios of SLH predictions to current limits with $t_\beta=1$ and $x_U=x_D=1$.}
\label{plot1}
\end{figure}

\begin{figure}[htp]
 \centering
  \includegraphics[height=6.1cm]{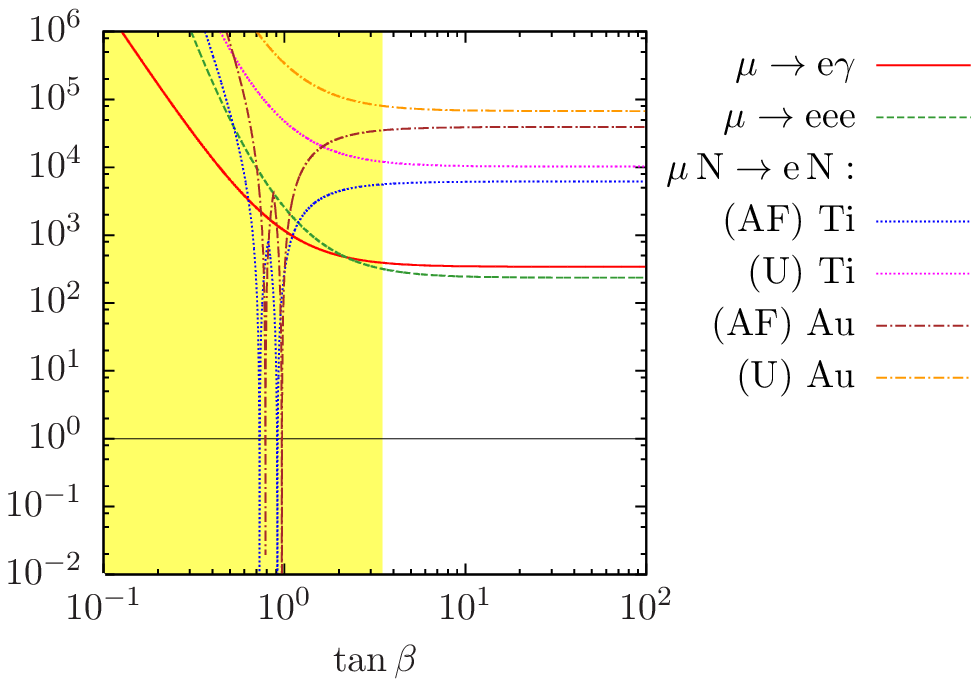}
\caption{Ratios of SLH predictions to current limits as functions of $t_\beta$ with natural values for all other parameters. The shaded region on the left is excluded by the limit on the mixing of SM leptons with heavy neutrino singlets.}
\label{tbeta}
\end{figure}

In figure \ref{plot1}, we plot the branching ratios normalized by their current experimental limits as functions of each of the parameters $f$, $\tilde{x}$, $\delta$ and $\theta$.  In each case, we vary one of the parameters and keep the remaining ones at the reference values.  Only normalized values below the unity are experimentally allowed.  The general behavior of the SLH model is very similar to that of the LHT model studied previously in \cite{delAguila:2008zu,delAguila:2010nv}.  Tables \ref{limits1} and \ref{limits2} contain generic constraints on the scale $f$, $\delta$ and $\sin{2\theta}$ for standard values for the rest of the parameters in each case.  
As can be observed in the $\tilde{x}$ plot, there are cancellations between different contributions to $\mueN$ conversion processes, but for rather low heavy neutrino masses.\footnote{In this example, the cancellations occur from interference among boxes and penguins. A similar effect can be seen in the LHT model (see figure 4 of Ref.~\cite{delAguila:2010nv}), where the $\gamma$ and $Z$ penguins can equate the box contribution in absolute value at specific points but with opposite sign. In the same figure, one can also see that the penguin contributions vanish at some points, where they flip sign. Although we only show cancellations in $\mueN$ conversion, they are not necessarily limited to this case and it is possible that different combinations of parameters could generate cancellations in other processes.}
The abrupt change of the slope along the curves away from zero reflects the change of the sign 
assignment $\epsilon$ in (\ref{deltaq}), which we choose to minimize the size of the 
conversion rate. 
Figure \ref{tbeta} shows the dependence on $t_\beta$.  Notice that, in general, this parameter does not allow us suppress the branching ratios enough to get within the experimental bound.  Furthermore, the dependence on $t_\beta$ is very light for larger values of this parameter due to the fact that it appears in the denominator of $\delta_\nu$ and simply suppresses these terms for large values.  Low values for $t_\beta$ makes the calculation unstable (the expansions of the scalar fields would no longer be good aproximations), being forbidden anyway because of the limit $\delta_\nu<0.05$.

\begin{figure}[htp]
 \includegraphics[width=\linewidth]{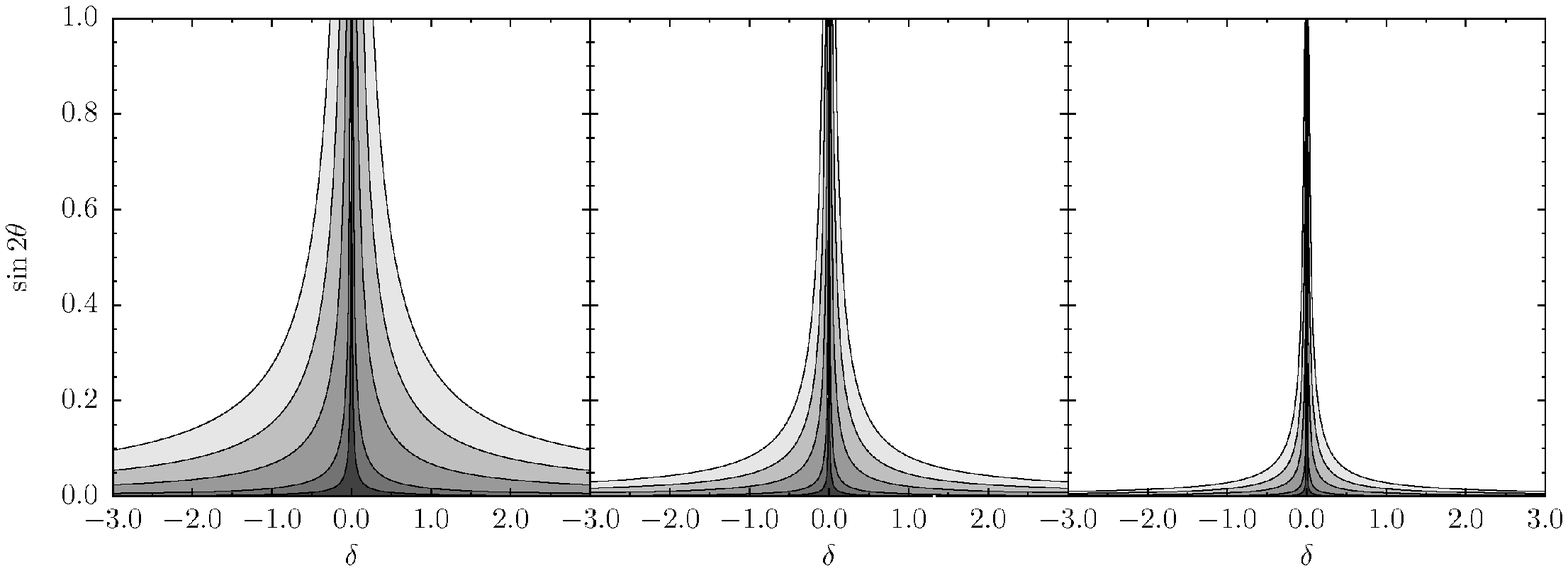}\\
\includegraphics[width=\linewidth]{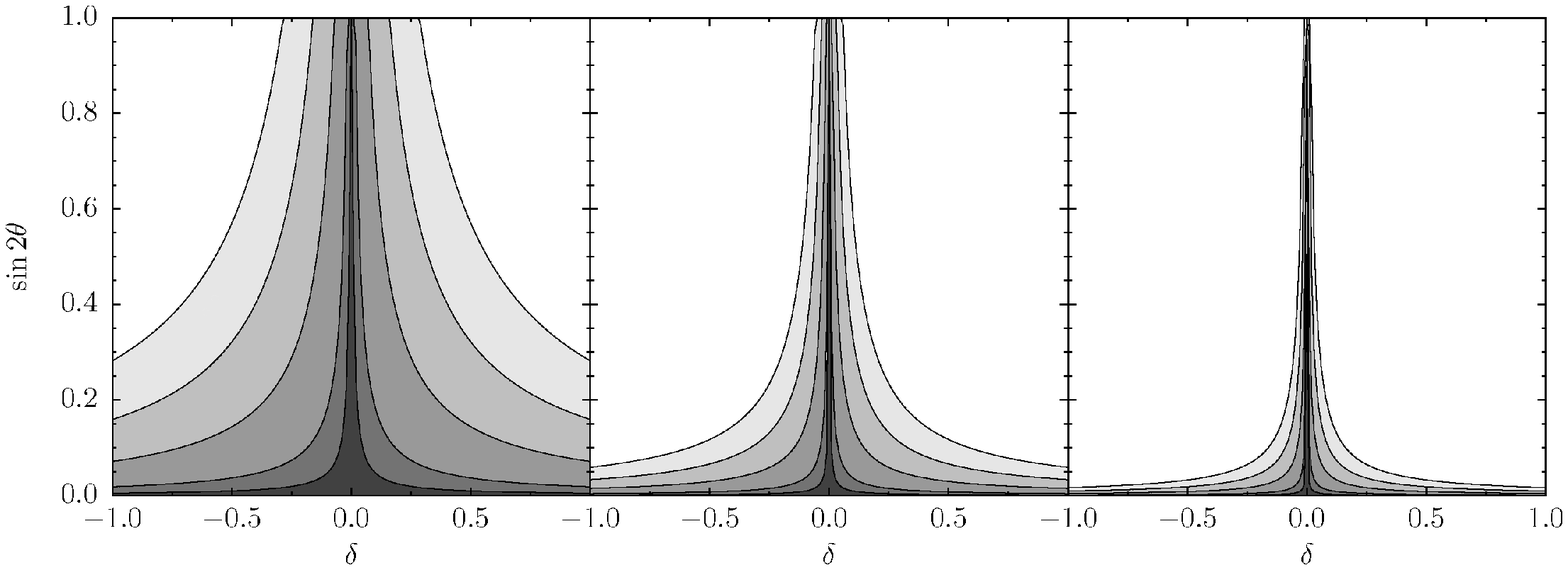}
\caption{Exclusion contours in the ($\delta$, $\sin 2\theta$) plane for the anomaly free embedding (top) and the universal embedding (bottom).  Each curve corresponds to a different value of $f$ (from the bottom up $f = 0.5$, 1, 2, 3, 4 TeV) and each plot to a different value of $\tilde{x}$ (from left to right $\tilde{x}=0.5$, 1, 4).  Also $x_U = x_D = 1$ and $t_\beta = 1$.}
\label{contour}
\end{figure}

In figure \ref{contour} we show exclusion contours for current experimental limits in the ($\delta$, $\sin 2\theta$) plane.  Points below the contour lines are within the measured bounds for all studied LFV processes although, in general, the $\mu\, {\rm Au} \to \e\, {\rm Au}$ process gives the most stringent limits.  These plots show that the mass splitting and the mixing angle must be correlated in order to suppress the LFV effects.  The correlation is similar for both embeddings, both of which require very small mixing angles or mass splittings to stay within experimental constraints.

\begin{figure}[htp]
\centering
\includegraphics[height=13cm]{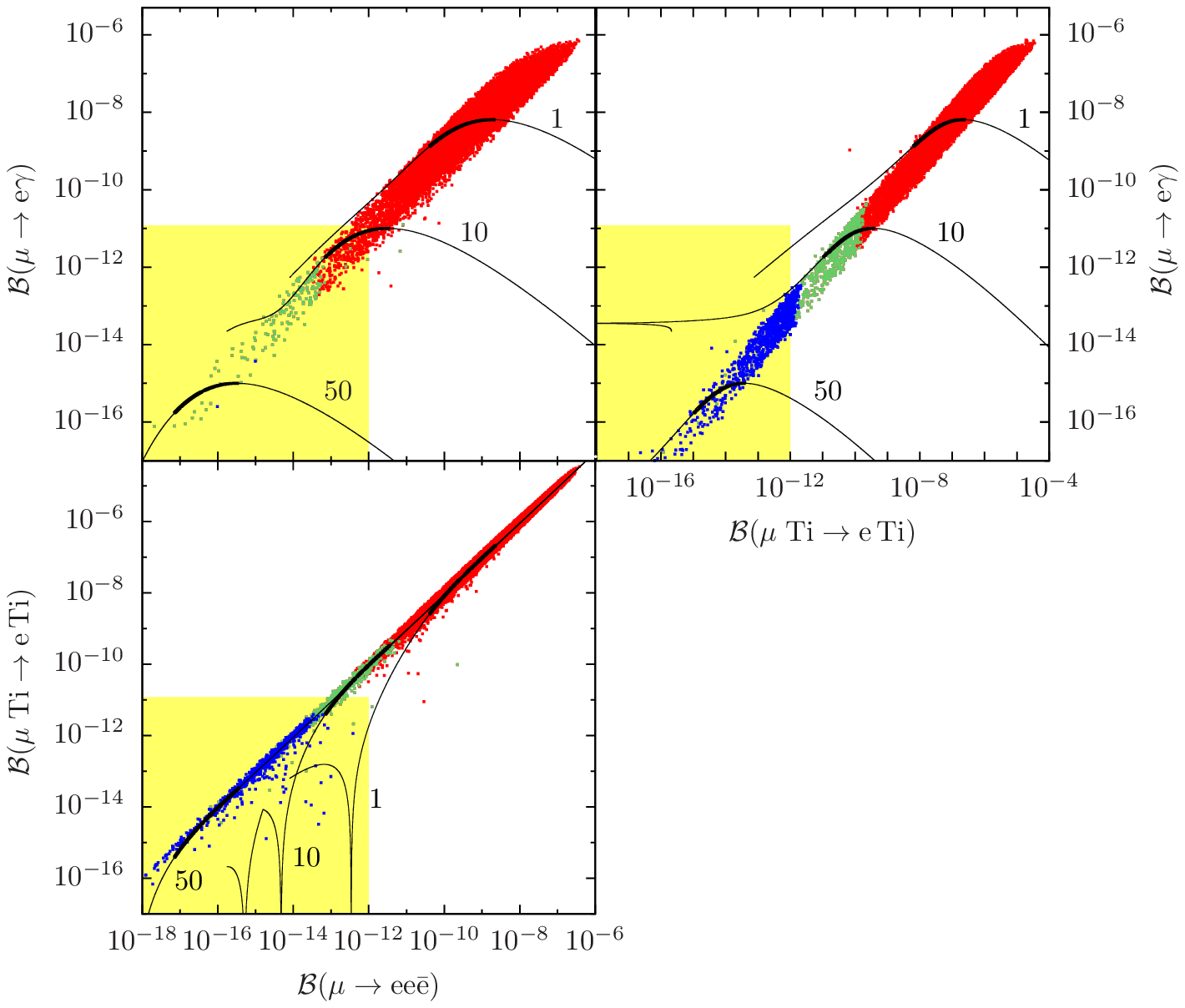}
\caption{Scatter plots for random scans over model parameters (see text).}
\label{scatter}
\end{figure}

In figure \ref{scatter} we show scatter plots for the different processes.  The points are generated from a random scan of the parameter space which was done taking $\tilde{x}$ uniform in the range $[0.25,4]$, $\delta$ uniform on the range $[-4,4]$, $f^{-4}$ uniform in a range corresponding to $f\in [0.5,50]$ TeV, $\sin{2\theta}$ uniform in $[0,1]$ and $t_\beta$ uniform in $[0.1,10]$.  We then discard points that do not comply with the limit $\delta_\nu < 0.05$, i.e., $ft_\beta > 3.48$ TeV to obtain the red points.  The green points are generated by also excluding points that do not fulfill the current limits on the process which is not plotted. For instance, if we are plotting $\muegamma$ against $\mueee$, we exclude (red) points that have the $\mueN$ process over its current experimental bound.  The blue points are analogous to the green points but we use the future limits on the complementary process rather than the current limit.  We show $\muegamma$, $\mueee$ and $\mueN$ on Ti (the future limit for this nucleus is the strongest)
in the anomaly free embedding (both embeddings produce similar results and we choose just one for simplicity).  The shaded area is allowed by current experimental limits on the plotted processes.  No blue points are plotted in the first graph because the future limit on $\mueN$ on Ti excludes practically all points in the scanned range.
In all cases, we see that the branching ratios are clearly correlated with each other due to the fact that the basic behavior is the same in all cases, as given in equation (\ref{dep}).  The only deviations from this come mainly from the dependence on $\tilde{x}$ and, to a lesser degree, from $t_\beta$ which is different in each case.  The range chosen for this parameter roughly fixes the width of the scatter plot while the ranges for the other parameters (notably $f$) fix the length of the plot.  
For illustration, we draw the curves of constant $f = 1, 10, 50$ TeV and varying $\tilde{x}$.  The thicker part of these lines correspond to the allowed $\tilde{x}$ range in the scatter plots $[0.25,4]$.  The edges of the point distribution approximately align with the edges of these thicker sections.  The alignment would be exact if the scaling (\ref{dep}) were exact and if there were no dependence on $t_\beta$.
In the cases where we plot $\mueN$ we see the effect of the cancellations from figure \ref{plot1} which 
translates into the thin lines approaching vanishing values of the conversion rate for this process.  
The point density is not representative of the relative frequency of each point type since they are obtained from separate runs (they are not subsets of a single set of points) and are forced to have a specific number of points each.  Also, different choices of the parameter distributions can produce other point densities.  The purpose of the plot is to illustrate the area reachable while keeping the parameters in the specified ranges.  This means that, to some extent, experimental detection of one of these processes constrains the possible values for the others.

\section{Conclusions}
\label{conclusions}

Although lepton flavor mixing in the SLH model had been previously addressed \cite{Han:2005ru}, no phenomenological studies of the basic flavor changing proceses had been presented before. To this end, a thorough analysis of the Lagrangian has been performed in order to obtain all the required field interactions in the 't Hooft-Feynman gauge, which is well suited for one-loop calculations. In particular, we had to identify to the desired order in $v/f$ the actual would-be-Goldstone bosons, those to be eaten by the corresponding physical gauge bosons. Then the Feynman rules for all needed interactions were obtained.

We have analytically calculated the amplitudes of the basic lepton flavor changing processes in terms of standard one-loop functions, that have been reduced to relatively simple expressions. These amplitudes are ultraviolet finite, as they are in the case of the LHT model \cite{delAguila:2008zu,delAguila:2010nv,Goto:2008fj,Blanke:2009am}. 
On the other hand, the heavy neutrino Yukawa exhibits a non-decoupling behavior (figure~\ref{plot1}): both $Z$ and $Z'$ penguin contributions to the amplitudes grow with $m^2_N/M_X^2$, a result that is well known and was discussed in \cite{delAguila:2008zu,delAguila:2010nv}.

To simplify the phenomenological analysis, we have assumed, as in previous studies, that only two lepton families mix. This leaves us with just four free parameters. Present limits on the considered rare processes translate into bounds on these parameters that allow us to assess the degree of naturalness of the model. From table \ref{limits1} we conclude that the value of the breaking scale $f$ must be above 14 TeV, which is more stringent than the limit derived from EWPD, about 4 TeV 
\cite{Csaki:2002qg,Csaki:2003si,Han:2005dz,Chen:2006dy}.\footnote{
This bound also applies to variations of the original SLH model 
\cite{Schmaltz:2004de} implementing a lighter new $T$ quark and a richer 
Higgs phenomenology consistent with present data \cite{Barcelo:2007if,Barcelo:2008je}. 
Such an extension does not alter the flavor structure discussed in this work.} 
Thus, near our standard reference values, flavor constraints in the SLH model and in the LHT model are similar.  For these specific values, the strongest limit on the SLH $f$ scale is of almost 14 TeV for the anomaly free embedding while the analogous limit in the LHT model is of about 10 TeV \cite{delAguila:2010nv}.  The universal embedding is similar with a limit of about 16 TeV.  However, we must keep in mind that these limits are not rigid and depend on the reference point. Moreover, the point was different for LHT than for SLH ($\tilde{x}$ was taken at 4 in the LHT in order to avoid an unnatural cancellation).  Additionally, for equal $f$, the gauge bosons in the LHT model are heavier than those in the SLH model by about a factor $\sqrt{2}$.  Keeping these differences in mind we find that the results for either model are similar and of the same order of magnitude.

As can be observed in the scatter plots in figure \ref{scatter}, 
present LFV experimental limits seem to allow a relatively large region of the parameter space (green and blue points in the shaded regions), 
as is the case for the LHT model \cite{Blanke:2007db,Blanke:2009am}.  
This does not contradict our main claim about the necessary fine tuning. 
Indeed, in order to be in the experimentally allowed area the effective 
misalignment between the light and heavy lepton flavors $\delta \sin 2\theta$, 
depending not only on the rotation angle but also on the heavy mass difference, must be very small and at the 
per cent level (or otherwise $f$ of the order of several TeV and relatively large). 
On the other hand, there is a large correlation in both models between 
$\muegamma$ and $\mueee$ predictions. This is so because both processes scale 
to a very good approximation as $|\frac{v^2}{f^2}\delta \sin 2\theta|^2$ 
(see equation \ref{dep}) and the dependence on the other parameters is mild 
within their natural range. 
The situation is similar for $\mue$ conversion because 
in the natural region there are no large cancellations among different contributions. 
However, in general there are narrow regions in parameter space where these 
processes can vanish as illustrated in figure \ref{plot1} and in figure \ref{scatter} by the black curves going to zero values of the 
conversion rate for this process. 
Obviously, there are many other parameters in a formulation with three families 
but the behavior will be analogous using a convenient parametrization for the 
misalignment \cite{delAguila:2008zu}. 
The former constraints significantly restrict model building, 
giving rise to a little flavor hierarchy problem also in LH scenarios. 
The fact that these models are so sensitive to present LFV processes, however, also implies that they could explain possible future observations of LFV by MEG \cite{Mori:2007zza,Adam:2009ci} at PSI or by PRISM/PRIME \cite{Kuno:2005mm,Pasternak:2010zz} at J-PARC.

Finally, although for simplicity we have neglected any flavor violation in the 
quark sector, this aspect deserves special attention. 
In particular, it may explain a possible discrepancy 
between the SM prediction and the measured value of 
quark rare processes, like for instance the muonic $B$ decays 
$B^0_s \rightarrow \mu \bar{\mu}$ and $B^0 \rightarrow K^*\mu \bar{\mu}$ 
\cite{Altmannshofer:2008dz} 
(see \cite{Buras:2009if} for a review and further 
references), 
to be precisely measured at LHCb 
\cite{Alves:2008zz,Tuning:1314837}.  
Present upper bounds on the former by the CDF \cite{:2007kv} 
and D0 \cite{Abazov:2010fs} Collaborations are still almost one order 
of magnitude above the SM prediction, 
whereas the asymmetry measurements of the 
latter by BABAR \cite{Aubert:2008ps}, BELLE \cite{:2009zv} and CDF \cite{Aaltonen:2008xf,Pn10047} 
may hide a hint of new physics.
At any rate, the quark sector of the SLH model allows for a quite rich phenomenology 
because in contrast with the LHT model, 
there are FCNC already at tree level, only suppressed by 
small mixing angles, and eventually of the same size as the one-loop effects 
resulting from the exchange of new particles only.  
As we have discussed, one can consider two different  
completions of the strong sector, one universal with an extra vector-like 
quark of charge $2/3$ per family and an anomaly free completion 
with an extra vector-like quark of charge $1/3$ for each of the 
first two families and an extra vector-like quark of charge $2/3$ 
for the third one. 
In the universal case there are no tree-level flavor changing 
$Z$ couplings in the down sector but there are in the up sector, 
whereas there are tree-level FCNC for both types of quarks 
in the anomaly free case \cite{delAguila:1982fs,delAguila:2000rc}. In either case there are 
one-loop FCNC also contributing to the quark transitions. 
Studying the corresponding 
flavor constraints would be also interesting 
in order to discriminate this model from other SM extensions \cite{Altmannshofer:2009ne,Buras:2010wr}. 
Let us note too that
although the LFV bounds may be a priori more stringent than 
those from quark flavor violation, the SLH 
contributions to charged lepton transitions are one-loop suppressed 
and the corresponding limits on some model parameters 
may be eventually comparable to those derived 
from hadronic tree-level processes.
   
\subsection*{Acknowledgments} 
Work supported by the Spanish MICINN (FPA2006-05294), Junta de Andaluc{\'\i}a (FQM 101, FQM 03048) and European Community's Marie-Curie Research Training Network under contract MRTN-CT-2006-035505 ``Tools and Precision Calculations for Physics Discoveries at Colliders''. M.D.J. was supported by a MICINN FPU fellowship.

\appendix
\section{Loop functions}
\label{loopfunctions}
The most general 3-point functions can be written as $C(p_1^2,Q^2,p_2^2; M_1^2,M_2^2,M_3^2)$.  In our case $p_1^2=p_2^2=0$ and only the following general types are relevant for our analysis: $C\equiv C(0,Q^2,0,M_1^2,M_2^2,M_2^2)$, $\overline{C}\equiv C(0,Q^2,0,M_2^2,M_1^2,M_1^2)$ and $\hat{C}\equiv C(0,Q^2,0,M_1^2,M_2^2,0)$ (the last one, symmetric under the exchange $M_1\leftrightarrow M_2$).  We define the mass ratio $x=M_2^2/M_1^2$ and reparameterize the functions in terms of $M_1^2$, $Q^2$ and $x$. The resulting functions are the same as those in \cite{delAguila:2008zu,delAguila:2010nv}.  We recall their expressions:
\begin{align}
C_0(M_1^2,Q^2;x)&=\frac{1}{M_1^2}\bigg[\frac{1-x+\ln x}{(1-x)^2}
\nn\\ 
&+\frac{Q^2}{M_1^2}\frac{-2-3x+6x^2-x^3-6x\ln x} {12x(1-x)^4}\bigg]+{\cal O}(Q^4),\quad
\\
\overline C_0(M_1^2,Q^2;x) &=\frac{1}{M_1^2}\bigg[\frac{-1+x-x\ln x}{(1-x)^2}
\nn\\
&+\frac{Q^2}{M_1^2}\frac{-1+6x-3x^2-2x^3+6x^2\ln x} {12(1-x)^4}\bigg]
+{\cal O}(Q^4),\quad
\\
C_1(M_1^2,Q^2;x)&=\frac{1}{M_1^2}\frac{-3+4x-x^2-2\ln x}{4(1-x)^3} +{\cal O}(Q^2),
\\
\overline C_1(M_1^2,Q^2;x) &=\frac{1}{M_1^2}\frac{1-4x+3x^2-2x^2\ln x}{4(1-x)^3} +{\cal O}(Q^2),
\\
C_{11}(M_1^2,Q^2;x)&=\frac{1}{M_1^2}\frac{11-18x+9x^2-2x^3+6\ln x}{18(1-x)^4}+{\cal O}(Q^2),
\\
\overline C_{11}(M_1^2,Q^2;x)&=\frac{1}{M_1^2}\frac{-2+9x-18x^2+11x^3-6x^3\ln x}{18(1-x)^4} +{\cal O}(Q^2),
\\
C_{00}(M_1^2,Q^2;x)&=-\frac{1}{2}B_1
-\frac{Q^2}{M_1^2}\frac{11-18x+9x^2-2x^3+6\ln x}{72(1-x)^4}
+{\cal O}(Q^4),
\\
\overline C_{00}(M_1^2,Q^2;x)&=-\frac{1}{2}\overline B_1
-\frac{Q^2}{M_1^2}\frac{-2+9x-18x^2+11x^3-6x^3\ln x}{72(1-x)^4}
+{\cal O}(Q^4),\nn \\
\hat{C}_{00}(M_1^2,Q^2;x) & = \frac{1}{8} \left( 3 + 2 \Delta_\epsilon - 2\ln \frac{M_1^2}{\mu^2} \right) + \frac{x\ln x}{4(1-x)} + {\cal O}(Q^2).
\end{align}

The 2-point functions are written in a similar way.  Their general form is $B(p^2;M_1^2,M_2^2)$, but we only need functions $B\equiv B(0;M_1^2,M_2^2)$ and $\overline{B} \equiv B(0;M_2^2,M_1^2)$.  Only $\overline{B}_1$ is necessary.
\begin{equation}
 \overline{B}_1(M_1^2;x) = -1 - \frac{1}{2}\left(\Delta_\epsilon - \ln{\frac{M_1^2}{\mu^2}} \right) + \frac{3-4x+x^2+2x^2 \ln{x}}{4(1-x)^2}.
\end{equation}

The 4-point functions are given by $D(p_1^2,p_2^2,p_3^2,p_4^2,(p_1+p_2)^2,(p_2+p_3)^2;M_0^2,M_1^2,M_2^2,M_3^2)$ and we need only functions with zero external momenta $D(0,0,0,0,0,0;M_0^2,M_1^2,M_2^2,M_3^2)$.  These functions are then symmetric under any exchange of masses.  We define $d_0 = M_0^4 D_0$ and $\tilde{d}_0 = 4 M_0^2 D_{00}$ so that these functions now depend only on 3 mass fractions: $M_{1,2,3}^2/M_0^2$ which we denote $x$, $y$ and $z$ respectively.  The general expressions required are as follows \cite{delAguila:2008zu,delAguila:2010nv}:
\begin{eqnarray}
d_0(x,y,z)&=&
\frac{x\ln x}{(1-x)(x-y)(x-z)}-\frac{y\ln y}{(1-y)(x-y)(y-z)}\\
&& +\frac{z\ln z}{(1-z)(x-z)(y-z)}, \\
\widetilde d_0(x,y,z)&=&
\frac{x^2\ln x}{(1-x)(x-y)(x-z)}-\frac{y^2\ln y}{(1-y)(x-y)(y-z)}\\
&& +\frac{z^2\ln z}{(1-z)(x-z)(y-z)}.
\end{eqnarray}

The case in which $z\rightarrow 1$ is also necessary:
\begin{eqnarray}
d_0(x,y)&=&
-\left[
\frac{x\ln x}{(1-x)^2(x-y)}-\frac{y\ln y}{(1-y)^2(x-y)}+\frac{1}{(1-x)(1-y)}
\right],
\\
\widetilde d_0(x,y)&=&
-\left[
\frac{x^2\ln x}{(1-x)^2(x-y)}-\frac{y^2\ln y}{(1-y)^2(x-y)}+\frac{1}{(1-x)(1-y)}
\right].
\end{eqnarray}


\bibliographystyle{JHEP}
\bibliography{paper3}{}

\end{document}